\documentclass[preprint,aps,prb]{revtex4}
\usepackage[dvips]{graphicx}

\def\Ntotal{N_\mathrm{total}}
\def\Nmicelle{N_\mathrm{micelle}}
\def\Nsmall{N_\mathrm{small}}
\def\theal{t_\mathrm{healing}}
\def\tgen{t_\mathrm{generation}}

\def\bh{\mathbf{h}}
\def\bk{\mathbf{K}}
\def\Peq{P_\mathrm{eq}}
\def\nrepeat{N_\mathrm{repeat}}
\def\ett{\epsilon_{TT}}
\def\eth{\epsilon_{TH}}
\def\ehh{\epsilon_{HH}}
\def\bx{\mathbf{x}}
\def\bbf{\mathbf{F}}
\def\bg{\mathbf{g}}
\def\by{\mathbf{y}}
\def\bej{\mathbf{e_j}}
\def\bxzero{\mathbf{x_0}}
\def\bxj{\mathbf{x_j}}
\def\Pgauss{P_\mathrm{Gauss}}

\begin{document}

\title{Coarse Grained Computations for a Micellar System.}
\author{Dmitry I. Kopelevich}
\altaffiliation[Current address: ]
            {Department of Chemical Engineering
	     University of Florida,
             Gainesville, FL 32611}
\email{dkopelevich@che.ufl.edu}
\author{Athanassios Z. Panagiotopoulos}
\email{azp@princeton.edu}
\author{Ioannis G. Kevrekidis}
\thanks{Corresponding author}
\email{yannis@arnold.princeton.edu}
\affiliation{Department of Chemical Engineering\\
Princeton University\\
Princeton, NJ 08544}
\begin{abstract}
We establish, through coarse-grained computation, a connection
between traditional, continuum numerical algorithms (initial value problems as well
as fixed point algorithms) and atomistic simulations of the Larson model
of micelle formation.
The procedure hinges on the (expected) evolution of a few slow, coarse-grained 
mesoscopic observables of the MC simulation, and on (computational) time scale separation
between these and the remaining ``slaved", fast variables.
Short bursts of appropriately initialized atomistic simulation are 
used to estimate the (coarse-grained, deterministic) local dynamics of the
evolution of the observables.
These estimates are then in turn used to accelerate the evolution to computational
stationarity  through traditional continuum algorithms (forward Euler integration, 
Newton-Raphson fixed point computation).
This ``equation-free" framework, bypassing the derivation of explicit, closed
equations for the observables (e.g. equations of state) 
may provide a computational bridge between direct
atomistic / stochastic simulation and the analysis of its macroscopic, system-level
consequences.

\end{abstract}
\maketitle

\section{Introduction}

Textbook reaction/transport modeling is based on macroscopic equations -
typically partial differential equations embodying 
conservation laws closed through constitutive relations.
The Navier-Stokes equations as a model of laminar fluid flow provide
a good ilustrative example: they describe the behavior of a very complex 
system (molecular collisions and interactions with the flow boundary) at a level of 
coarse-graining (velocity and pressure fields) which is practical for engineering
design (e.g. pressure drop computations).
What makes this possible is {\it closures}: modeling the stresses as functionals of
the velocity field through Newton's law and viscosity.
%
%
In engineering modeling we increasingly encounter systems whose
coarse-grained, mesoscopic behavior emerges from the interaction of a number of
``agents" (molecules in a fluid, cells in a tissue, individuals in a population)
between themselves and with their environment. 
It is the coarse-grained behavior that we want to predict, design or control; yet
the available models are only available at a much finer, atomistic or stochastic
level.
The closures that will translate these models to ``practically predictive"
mesoscopic level models are simply not available in closed form.

Over the last few years we have been developing an ``equation free" 
computational methodology for extracting coarse-grained 
information from microscopic models; this methodology
provides an alternative to direct, long-term atomistic/stochastic simulation.
Our goal is to accelerate the computation of coarse-grained quantities
or properties by acting directly, and as parsimoniously as possible,
on the direct, full scale (kinetic Monte Carlo, kMC, molecular
dynamics, MD, Brownian dynamics, BD, also quantum chemistry based
simulators like such as Car-Parrinello MD, CPMD) simulator.
The basic idea is to use the microscopic simulator as a computational
experiment that we can initialize at will.
Short bursts of appropriately initialized atomistic simulations are 
designed, executed, and their results processed to provide ``on demand"
the information that one would obtain from coarse-grained models,
had these models been available in closed form.
This system identification based ``closure on demand" approach, provides
a bridge between traditional continuum numerical analysis (integration, 
steady state computation, the solution of linear and nonlinear equations, optimization) and
atomistic / stochastic / individual based simulators.
Short, appropriately initialized dynamic simulations of the detailed model 
(``coarse timestepping") enable the microscopic simulator to perform tasks
(like locating, quantifying the stability of, or optimizing 
coarse grained stationary states)
that it has not been explicitly designed for.
What makes it possible is the ability of a modeler to {\it initialize a
computational experiment at will}; initializing a laboratory experiment
at will is an immensely more difficult task.

This coarse-grained, equation-free computational framework
was introduced in Ref.~\onlinecite{Theodoropoulos00}, and has since been used
in a number of modeling contexts (from coarse Brownian dynamics of
nematic liquid crystals \cite{SGK03} to coarse molecular dynamics for alanine
dipeptide folding \cite{HK03}, 
individual-based modeling of evolving diseases \cite{CGLK04} and
lattice gas modeling of surface reactions \cite{MMK02,MMPK02}), see the reviews in 
Refs.~\onlinecite{Manifesto,ShortManifesto}.

The main assumption in this framework is that a separation of time
scales (and concomitant space scales) prevails in the description of
the system state; indeed, that we can separate the state in a subset of 
slow (coarse, macroscopic) variables $\bx$ and its complement - a subset
of fast variables $\by$. 
Typically, the slow variables consist of a few of the lower moments
of microscopically / stochastically evolving distributions (such as
concentration for chemical reactions,  or density and momentum, the zeroth
and first moments of a distribution of flowing molecules over velocities).
Over the time scale of interest, the dynamics of the fast variables 
become quickly slaved to the dynamics of the slow, ``master" variables. 
If this assumption holds (and this is a reasonable assumption if the 
system is believed to have reproducible coarse-grained behavior), 
then one does not need to derive the macroscopic equations explicitly.
One can in effect solve them, and perform system-level computational modeling
and design tasks with them without deriving them in closed form;
model evaluations are substituted by
``on demand" information from appropriately initialized microscopic simulations. 

The main tool that allows the performance of numerical tasks at the macroscopic level
using microscopic simulation codes is the so-called ``coarse timestepper"
\cite{Theodoropoulos00}.
The timestepper (which we denote by $\Phi_\tau$) is an operator 
which maps the state variables $\bx(t)$ forward by time $\tau$,
i.e. $\bx(t + \tau) =  \Phi_\tau\bx(t)$. 
If closed equations
for the evolution of the macroscopic variables $\bx(t)$ are 
available, then the timestepper is simply the
solution operator for these equations for time $\tau$.
When such equations are not available, the timestepper
can be estimated by the 
\emph{coarse timestepper}, which we denote by $\Phi^c_\tau$.
The coarse timestepper uses detailed microscopic
simulations of the system, and then processes the data
to estimate the time-$\tau$ map for the evolution of the
macroscopic observables for the problem of interest. 
The coarse timestepper consists of the following
basic elements (see Ref.~\onlinecite{Manifesto}):
\begin{enumerate}
\item A restriction operator, $M$, from the microscopic
      level description $\by$, to the macroscopic description, $\bx = M\by$.
      This operator usually involves averaging over microscopic space and
      over realizations of the ensemble of microscopic simulations.
\item A lifting operator, $\mu$, which constructs microscopic
      descriptions $\by$ consistent with the prescribed  macroscopic 
      description, $\bx$. This is obviously a ``one-to-many" operation.
      Lifting from the macroscopic to the microscopic description 
      and then restricting (projecting) again should have no effect, i.e. $M\mu = I$.
\item Prescribe a macroscopic initial condition $\bx(t = 0)$.
\item Transform it through lifting to an ensemble of consistent 
      microscopic realizations, $\by(t = 0) = \mu\bx(t = 0)$.
\item Evolve these realizations using the microscopic simulator
      for the desired short macroscopic time $\tau$, generating the values 
      $\by(\tau)$.
\item Obtain the restriction $\bx(\tau) = M\by(\tau)$ 
      and define the coarse timestepper as
      $\bx(t=\tau) = \Phi^c_\tau \bx(t = 0)$. 
      In other words, $\Phi^c_\tau = M\Phi_\tau \mu$.
\end{enumerate}

The use of such a coarse timestepper is based on (i) the separation of 
timescales between the micro- and the macroscopic descriptions of the 
system evolution, similar to that used in deriving explicit closed
macroscopic equations and (ii) on the assumption of existence of
a closed evolution equation {\it for the macroscopic variables} $\bx(t)$.
The basic premise is that the long-term dynamics of the microscopic 
simulation can be thought of as lying on a slow, low-dimensional
manifold which is {\it parametrized} by the coarse variables.
On this manifold, the remaining observables are slaved to (are
functionals of) the chosen coarse variables. 
Therefore, judicious choice of the coarse variables is crucial
for the correct implementation of coarse timestepping.
Coarse variables typically include a few lower moments of the 
microscopically evolving distributions, but any appropriate
order parameter set can be used as the observables of choice.
Once the coarse variables and the lifting and restriction operators are
constructed, the coarse timestepper procedure is easy
to realize numerically. 
It is also easy to parallelize: each ``lifted"
copy for the same coarse initial condition can be executed independently on 
a different processor over the same wall-clock time. 

Various system-level tasks can be implemented as 
computational superstructures around repeated calls to the
timestepper. 
In particular, the coarse projective Euler 
method \cite{Manifesto} is based on the extrapolation 
(projection) of the evolution of $\bx(t)$, obtained
through the application of the coarse timestepper,
over some macroscopic time step $\Delta t$.
A fixed point for the timestepper, i.e. a point
$\bx_0$ such that $\Phi^c_\tau(\bx_0) = \bx_0$, corresponds to an
equilibrium configuration of a thermodynamic system; this allows one
to consider using contraction mapping approaches,
like the Newton-Raphson method to obtain the equilibrium states.
In this work, least squares estimation is used to fit the
evolution of the macroscopic observables (as opposed to the simple
extrapolation in Ref.~\onlinecite{GK03a}) and to implement fixed point computations.
Other system-level tools that have been implemented using the
timestepper for various systems 
include bifurcation and stability analysis, coarse control,
coarse optimization, and the computation of coarse self-similar
solutions (Ref.~\onlinecite{Manifesto,ShortManifesto} and references therein.)

It is important to note that ``time" and ``evolution" in the 
above discussion do not have to be in real physical time; it could
be in the wall-clock time (iteration count) of an MC simulation.
In this paper we attempt to accelerate the convergence of
a Monte Carlo (MC) simulation to its ultimate stationary state.
We do not study how a real system dynamically approaches 
equilibrium; we only demand that our simulation faithfully reproduces the 
equilibrium, and try to accelerate the approach {\it of our
simulation} to {\it its} equilibrium.
It may be helpful to draw an analogy between physical evolution to 
a steady state, and the evolution of a Newton-Raphson iteration towards
the same steady state for {\it deterministic} problems.
The dynamics of the Newton-Raphson and the dynamics of physical evolution
are different, yet they share the same fixed point.
Our goal here is to accelerate (using coarse integration and coarse
fixed point techniques) the convergence of {\it our MC computation scheme}
to {\it its} equilibrium, which is hopefully shared with the real, dynamical 
problem (with ``nature").

The problem we will study using these coarse-grained, equation free methods,
is the spontaneous self-assembly of surfactant molecules in micelles;
we attempt to accelerate the (artificial) dynamics of an MC simulation of this
process.

It is well known that surfactant molecules, (amphiphiles,
i.e. molecules that contain both hydrophobic and hydrophilic
segments), spontaneously self-assemble into a variety of structures 
(see, e.g. Ref.~\onlinecite{Israelachvili_book}). 
At low amphiphile concentration, above the critical micelle concentration,
the surfactants assemble into micelles.
We study the micellization process using grand canonical
Monte Carlo (GCMC) simulations of the lattice model 
originally proposed by Larson \cite{LarsonMC85,LarsonMC96}. 
Despite its simplicity, this model leads to predictions 
(in particular, phase diagrams) that 
agree qualitatively with experiments \cite{LarsonMC96}.

Recently, Panagiotopoulos and coworkers 
\cite{AZPmicelles99,AZPmicelles02a} have studied micellization
using GCMC simulations of the Larson model.
In order to extrapolate the results to temperature and chemical
potentials different from those of actual runs, they used 
the histogram reweighting method
\cite{hist_reweighting1,hist_reweighting2}. 
This method
allows one to compute equilibrium distributions of the number
of molecules in the system $\Ntotal$ and the energy of the 
system $E$ at some temperature $T'$ and chemical potential
$\mu^{\prime}$ from the simulations 
performed at some other (nearby) $T$ and $\mu$. 
This method will not, however, extrapolate \emph{structural} properties,
such as micelle size or density;
coarse computational techniques (and, in particular, Newton-Raphson based continuation) 
might be a promising alternative in efficiently 
exploring the dependence on such variables on parameters.

This paper is organized as follows: In section II we discuss the
Larson model implemented in our MC simulation. In the two subsequent
sections we discuss our selection of coarse variables (Section III) and our
lifting procedure (the construction of microscopic states
consistent with coarse observables) (Section IV).
We then discuss and illustrate coarse projective integration (Section V)
and the coarse Newton method (Section VI).
In Section VII we discuss a kMC model fitted from our MC simulations,
and apply the same coarse-grained computational methods to it.
We summarize our results and conclude with a discussion in the 
final section (VIII).

\section{The Model}
In the Larson model \cite{LarsonMC85,LarsonMC96} studied here, 
three-dimensional space is discretized into a cubic lattice. 
An amphiphile molecule is modeled as a chain of 
beads and a solvent molecule is modeled by a single 
bead. 
The beads occupy sites on the lattice and the connected
beads of an amphiphile molecule are located on the
nearest-neighbor sites located along vectors
$(0,0,1)$, $(0,1,1)$, $(1,1,1)$ and their reflections along
the principal axis, resulting in a coordination number of 26.
There are two types of beads: hydrophobic tail (T) and hydrophilic
head (H). The solvent beads are assumed to be identical to the head beads.

Hydrophobic interaction is modeled by attractive interaction
between the tail beads. 
Each bead interacts only with its 26 nearest neighbors and
the total energy of the system is the sum of pairwise 
interactions between beads.
The tail-tail
interaction energy $\ett$ is -2 and the tail-head 
and head-head interaction energies $\eth$ and $\ehh$ are zero,
following Ref.~\onlinecite{AZPmicelles02a}.
It is furthermore assumed that all sites that are not occupied
by the amphiphile beads are occupied by the solvent. 
This latter assumption implies that there is no need to 
explicitly consider solvent in the Monte Carlo (MC) moves.

We perform grand canonical Monte Carlo (GCMC) simulations of this
model
and use the following mix of MC moves:
50\% amphiphile transfers (i.e. addition or removal), 
49.5\% amphiphile partial regrowth moves, and 0.5\% cluster moves.
We perform simulations of
a linear amphiphile chain H$_4$T$_4$ (which consists of 4 head beads 
and 4 tail beads) in a 
cubic box with the side length of 40 sites, assuming
periodic boundary conditions. 
This simulation box size is sufficiently large to prevent spurious 
effects of periodicity: a typical diameter of a micelle observed in the
simulations reported here is significantly
smaller than half the size of the box side.

\begin{figure}
\includegraphics[width=5in]{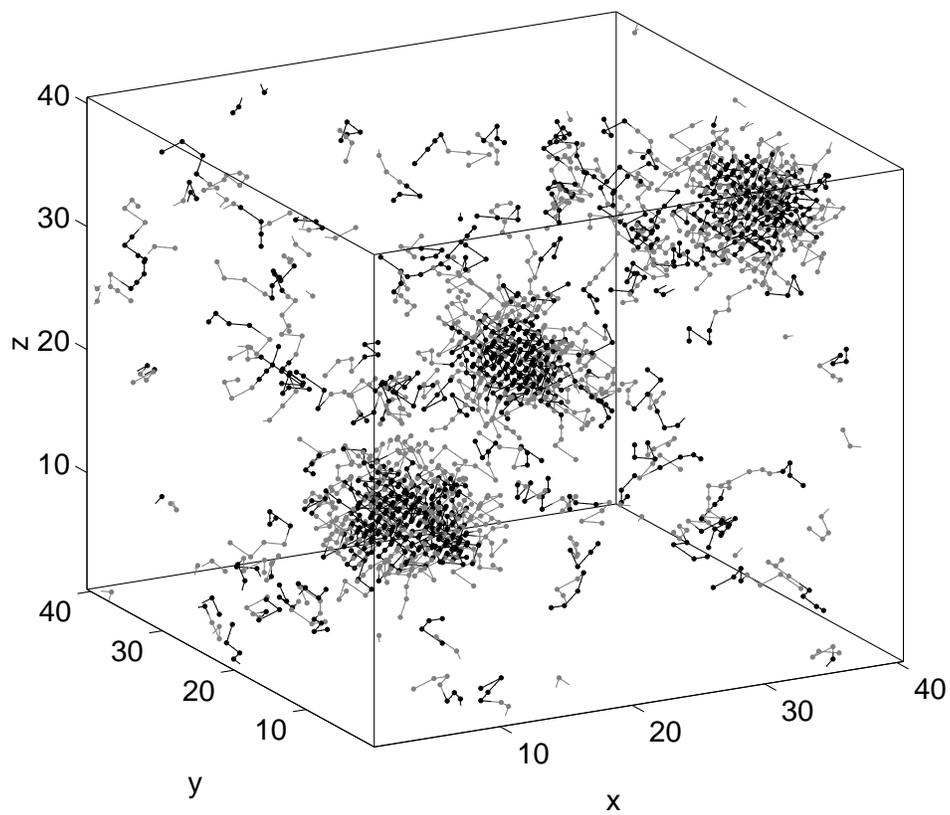}
\caption{Snapshot of a micellar system for temperature $T = 7.0$ and 
         chemical potential $\mu = -47.40$. 
	 Hydrophobic tail beads are shown in black and
	 hydrophilic head beads are shown in gray}
\label{F:snapshot}
\end{figure}

A snapshot of a simulation is shown in Fig.~\ref{F:snapshot}. 
In this 
example, we observe that the surfactant molecules have formed three micelles.
In addition, there is a significant fraction of smaller clusters. 
Our working definition of a cluster is an aggregate of amphiphile molecules
such that each molecule in a cluster has at least one tail bead which occupies
a neighboring site with a tail bead from another amphiphile of the cluster.
In other words, each cluster molecule interacts through hydrophobic attraction
with at least one other cluster molecule. 
The cluster size is defined
as the number of amphiphiles in this cluster. 
Note that
an isolated amphiphile molecule can be viewed as a cluster of 
size 1.

An example of the cluster size distribution obtained from the
GCMC simulations is shown in Fig.~\ref{F:coarse_vars}a and is
typical for a micellar system. 
There are two peaks in the distribution -- one
peak corresponding to the small clusters and the other peak corresponding 
to micelles. 
Note that there is an almost vanishing probability to 
observe a cluster of an intermediate size. 
This observation is
very important and allows us to define a set of coarse variables
(or ``coarse observables").

\section{Coarse Variables}
\label{S:coarse_var}

Choosing an appropriate set of coarse variables (``observables")
is an important step in the implementation of equation-free computation.
This choice is system-specific and should be
guided by physical intuition about the system,
or by data analysis techniques (e.g. Ref.~\onlinecite{PCA,SparseKernel}).
One of the requirements for the coarse variables is that they provide 
sufficient information about the system so that the ``lifting" operation
(micro from macro) can be successfully performed.
In earlier work on coarse computation \cite{MMK02,MMPK02,SGK03}, 
coarse variables were chosen to be moments of an evolving distribution.
Lifting then consisted of generating random realizations of a system 
configuration such that the average and, possibly, 
the variance of this quantity agreed with the prescribed values. 
More sophisticated lifting techniques (including short, constrained simulations in 
the spirit of algorithms like SHAKE in molecular dynamics \cite{SHAKE}) are also
being developed and tried.
Simple lifting approaches based on only a couple of moments
may not be satisfactory for more complex systems
such as the one studied in the current work. 
In addition to the averages,
we also need to preserve more of the structure of the system during the
transfer of information between micro and macro (lifting and restricting). 
More sophisticated approaches, which 
take into account the spatial inhomogeneity of a system, have
been presented by Gear \emph{et al.} \cite{Theodoropoulos02}.
In this so called micro-Galerkin method, an expansion
of the spatial distribution of the quantity of interest in a set of
global basis functions was used; these can be traditional polynomial
basis functions \cite{GearDistr} or possibly empirical eigenfunctions 
obtained by principal component data analysis from a microscopic simulation. 

%
%
Principal components of the raw data would give us some averaged structural 
information about a micelle, e.g. a density profile. 
It is not, however, trivial to reconstruct a micelle from its density profile. 
We believe that nonlinear data analysis techniques \cite{PCA,SparseKernel} hold the key
to systematic, non-intuitive choices of appropriate coarse observables.

In the case of our micellar system we kept
a database of cluster structures, and used as coarse observables
a number of features of the distribution of such structures; we
will describe these observables below.
In the lifting procedure, we place cluster
structures from the database directly into the simulation box 
according to certain prescribed distribution features.
We envision that for more complex self-assembled systems, such as 
bilayer or vesicles, one can combine the
empirical eigenfunction method with maintaining a structure database.

\begin{figure} 
\includegraphics[width=5in]{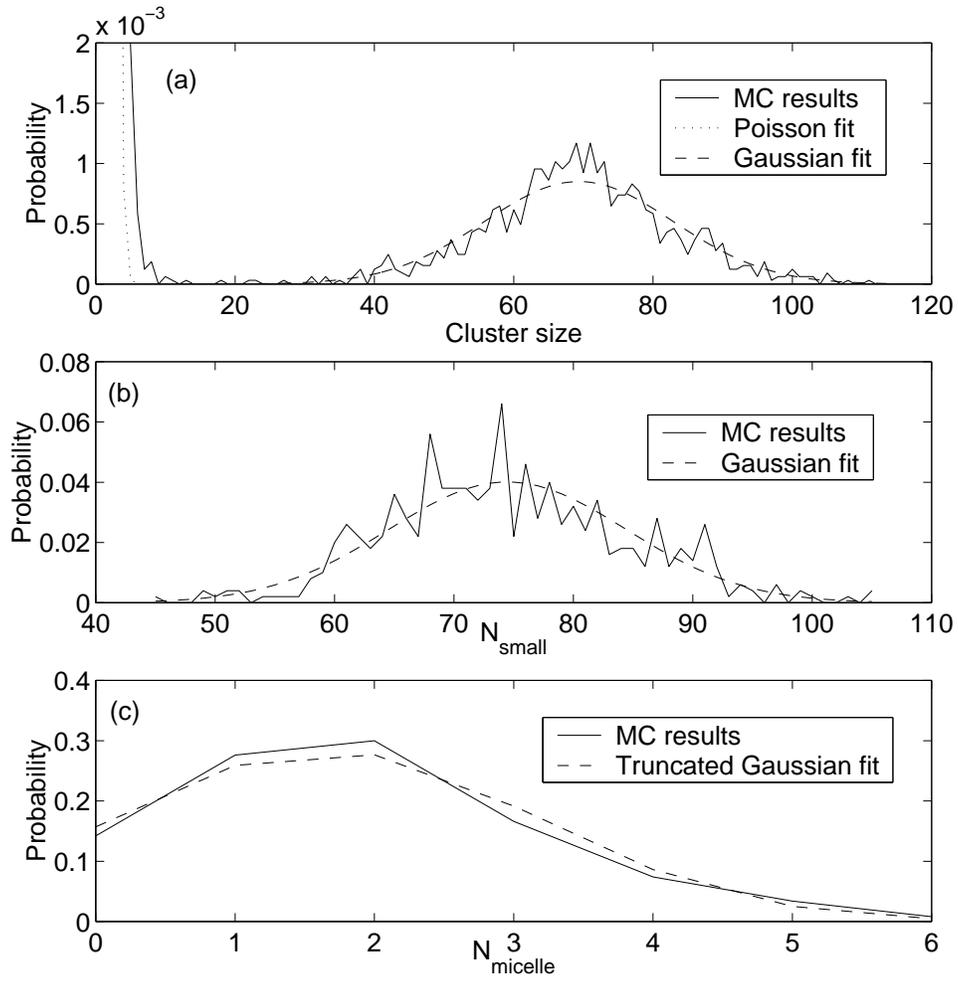} 
\caption{Definition of coarse variables: 
(a) cluster size distribution,
(b) distribution of number of molecules $\Nsmall$ contained in small 
    clusters, and
(c) distribution of number $\Nmicelle$ of micelles. 
    Distributions shown in this plot are obtained from averaging
    of 500 MC realizations.}
\label{F:coarse_vars}
\end{figure}

Details on computing the structure database will be provided in section~\ref{S:Euler}.
In addition to the database, we need to specify distributions of 
several quantities.
Describing these distributions through a finite number of macroscopic
observables (coarse variables) will be the basis of our coarse timestepper.
To choose a reasonable set of observables, capable of parametrizing
the evolution of our simulations, we examine representative simulation runs. 
Clearly, one needs to know the distribution of clusters.
In this work, we assume that the only quantity needed to characterize a
cluster is the number of amphiphiles. 
That is, all other physical properties
of the cluster (such as its radii of gyration and its energy) are quickly slaved
to the cluster size.
If we monitor clusters of a certain size, we observe
that over very short periods of simulation time, their physical attributes 
will ``approach a slow manifold" -- that is, their statistics will very
quickly become functions of the size. 
The analysis in the companion paper \cite{KPK04b} shows that
this is a reasonable assumption for the system considered here.
Hence, in order to characterize the clusters contained in the system,
we only specify their size distribution shown in
Fig.~\ref{F:coarse_vars}a.

In addition to the distribution of cluster sizes, we need to specify the 
concentration of amphiphiles in the system. 
Since we are performing simulations in the grand canonical
ensemble, in an equilibrium state, this concentration will fluctuate
around some average value. 
In principle, for a system of fixed volume, to specify concentration,
one can specify
the total number $\Ntotal$ of amphiphile molecules in 
the system. 
However, we found it more convenient
to ``split" this $\Ntotal$ into the ``small cluster" and the ``micellar 
components". 
This allows us to monitor the density of micelles
directly from our coarse variables and
to take advantage of the
separation of time scales between the micelles and the small clusters,
which is discussed below.
We furthermore assume that there is no correlation between sizes of 
different clusters in the system.
This assumption is reasonable because there are no long-range interactions 
in the system and the solution non-idealities are only due to excluded 
volume interactions. 
The systems considered here are dilute
and hence the non-idealities (correlations) are 
expected to be small. 
We will show it later that this assumption indeed holds.

To summarize, we suggest that the system can be
macroscopically successfully characterized by 
distributions of the following three quantities 
(see Fig.~\ref{F:coarse_vars}):
(i) cluster size, (ii) number $\Nmicelle$ of micelles in the system,
and (iii) number $\Nsmall$ of molecules contained in small clusters. 

A typical distribution of cluster sizes is shown in Fig.~\ref{F:coarse_vars}a. 
Clearly, the clusters
can be divided into two classes -- small clusters and micelles. 
Small clusters are unstable aggregates of a small number of molecules,
usually less than 10. 
They are being formed and destroyed 
very quickly during a simulation. 
For definiteness, we classify a cluster as a micelle if its 
size exceeds 20 molecules and as a small cluster otherwise. 
Since
there is essentially zero concentration of clusters of size between 20 and 40,
a precise location of the border between micelles and small clusters
is unimportant.

The cluster size distribution is further approximated by a Gaussian 
distribution (dashed line in Fig.~\ref{F:coarse_vars}a) 
and the small cluster size distribution is approximated 
by a Poisson distribution
(dotted line in Fig.~\ref{F:coarse_vars}a),
\begin{equation}
\label{eq:Poisson}
   P(k) = \frac{\lambda^{k-1}}{(k-1)!}e^{-\lambda},
\end{equation}
where $k = 1, 2, 3, \dots$ is the cluster size and $(\lambda + 1)$ is
the average size of a small cluster.
The size distribution of small clusters shows good agreement with 
the Poisson distribution because
the probability of formation of a small cluster consisting of $k$ molecules 
(in an ideal solution) is proportional to $\lambda^k$.

Hence, the cluster size distribution is described by three parameters:
average $\nu$ and standard deviation $\sigma$ of the micelle size and the
parameter $\lambda$ of the Poisson distribution of the small cluster 
sizes.
One more parameter is needed to completely specify
the cluster size distribution -- namely, the 
ratio between numbers of molecules contained in the small clusters and
in the micelles. 
This information is contained in the distributions of 
$\Nsmall$ and $\Nmicelle$ and the section~\ref{S:lifting} describes
the details of reconstruction of the system from these parameters.

Typical examples of distributions of $\Nsmall$ and $\Nmicelle$
are shown in 
Fig.~\ref{F:coarse_vars}b and ~\ref{F:coarse_vars}c, respectively.
The distribution of number $\Nsmall$ of molecules contained in small 
clusters is approximated by a Gaussian distribution
\begin{equation}
   P(\Nsmall) = \Pgauss(\Nsmall;\nu_s,\sigma_s).
\end{equation}
Here, $\nu_s$ and $\sigma_s$ are the mean and the standard deviation
of $\Nsmall$ and 
\begin{equation}
\label{eq:Gauss}
    \Pgauss(x;\nu,\sigma) = \frac{1}{\sqrt{2\pi\sigma}}\ 
      \exp(-(x-\nu)^2/2\sigma^2)
\end{equation}
is the Gaussian distribution with the mean $\nu$ and the standard 
deviation $\sigma$. As seen in Fig.~\ref{F:coarse_vars}b, the 
Gaussian distribution provides a good approximation to $P(\Nsmall)$ and 
hence in our coarse simulation, the distribution of $\Nsmall$ is 
completely specified by its first two moments, $\nu_m$ and $\sigma_s$.

The distribution of the number of micelles $\Nmicelle$ is 
approximated by a Gaussian truncated at $\Nmicelle=0$, i.e.
\begin{eqnarray}
\label{eq:trunc}
   P(\Nmicelle) & = & C \Pgauss(\Nmicelle;\hat{\nu}_m,\hat{\sigma}_m),
   \textrm{ if } \Nmicelle \ge 0, \\
   P(\Nmicelle) & = & 0, \textrm{ if } \Nmicelle < 0.
\end{eqnarray}
Here, $C$ is a normalization constant.
This truncation is important,
because there is a non-vanishing probability to have
zero micelles in the simulation box (see Fig.~\ref{F:coarse_vars}c).
If we were to fit an untruncated Gaussian to such a distribution, we would
end up with a non-vanishing probability to have a negative number of
micelles in the simulation box.
Note that, unlike the standard Gaussian, the parameters $\hat{\nu}_m$
and $\hat{\sigma}_m$ of $\Pgauss$ in Eq. (\ref{eq:trunc}) do not 
coincide with the mean $\nu_m$ and and the standard deviation
$\sigma_m$ of $\Nmicelle$. 
It is therefore necessary to obtain
these parameters 
from the requirement that 
the mean and the standard deviation of the truncated distribution
coincide with the mean $\nu_m$ and the standard deviation $\sigma_m$ of
$\Nmicelle$. 
This is done via
solution of a system of two nonlinear equations: 
\begin{equation}
   \nu_m = \sum P(\Nmicelle) \Nmicelle, \qquad
   \sigma_m^2 = \sum P(\Nmicelle) \Nmicelle^2 - \nu_m^2.
\end{equation}
Note that these
equations have a solution for only a subset of values of $\nu_m$ and
$\sigma_m$.
Fig.~\ref{F:coarse_vars}c shows that the truncated Gaussian provides 
a good approximation for $P(\Nmicelle)$.
Although we modified the standard Gaussian distribution
to fit $P(\Nmicelle)$, the distribution of the number of micelles
is still specified by its first two moments $\nu_m$ and $\sigma_m$.

To summarize, we chose to model the system by 7 coarse variables (observables):
3 for small clusters ($\lambda$, $\nu_s$, $\sigma_s$) and 
4 for micelles ($\nu$, $\sigma$, $\nu_m$, $\sigma_m$).

\begin{figure}
\includegraphics[width=5in]{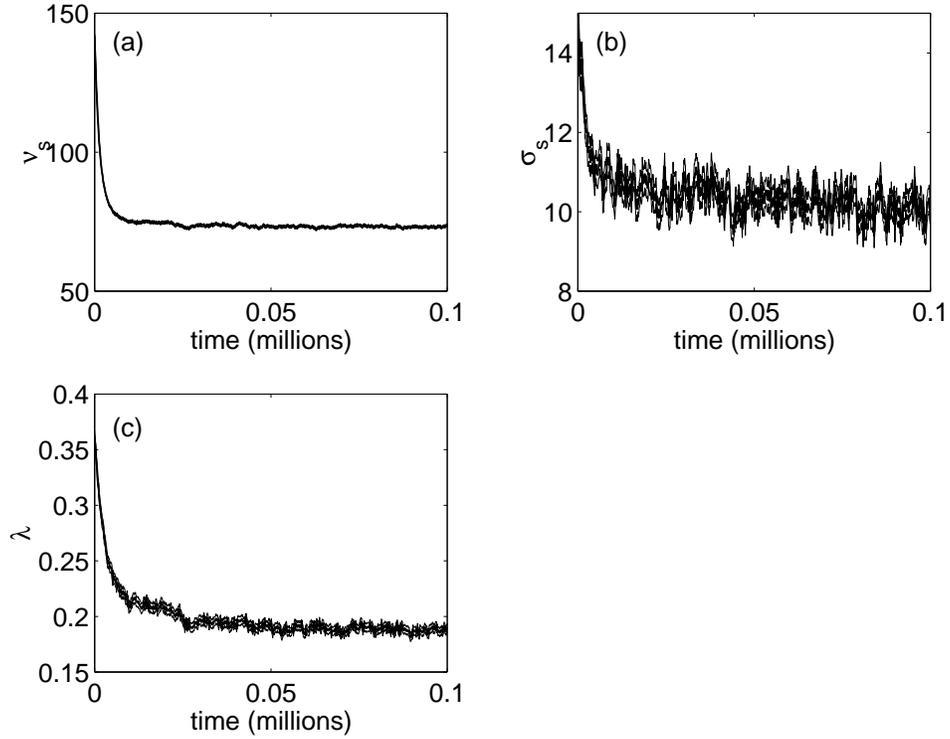}
\caption{Evolution of the small cluster parameters as the temperature
         and the chemical potential are switched from 
	 $T = 7.5$, $\mu = -46.20$ to $T = 7.0$, 
	 $\mu =-47.40$: (a) average small cluster size $\nu_s$,
	 (b) standard deviation $\sigma_s$ of the small cluster size, and
	 (c) average number $\lambda$ 
	 of molecules contained in the small clusters
	 The plots are obtained from averaging of 500 MC realizations.
         The error bars are indicated by the thin lines.
	 } 
\label{F:small_evol}
\end{figure}

\begin{figure}
\includegraphics[width=5in]{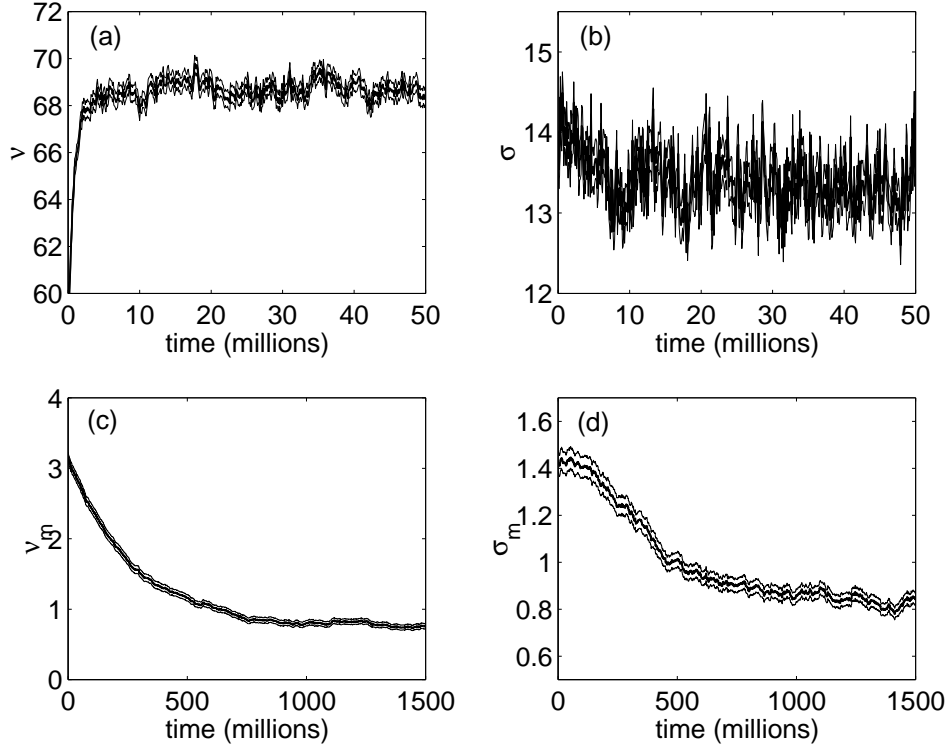}
\caption{Evolution of the micelle parameters as the temperature
         and the chemical potential are switched from 
	 $T = 7.5$, $\mu = -46.20$ to $T = 7.0$, 
	 $\mu =-47.40$: (a) average micelle size $\nu$, (b) standard
	 deviation $\sigma$ of the micelle size, (c) average number 
	 $\nu_m$ of 
	 micelles in the system, and (d) standard deviation
	 $\sigma_m$ of number of micelles in the system.
	 The plots are obtained from averaging of 500 MC realizations.
         The error bars are indicated by the thin lines.
	 }
\label{F:micelle_evol}
\end{figure}

Timescales (in MC iteration ``time") for the evolution of
these coarse variables are illustrated in 
Figures \ref{F:small_evol} and \ref{F:micelle_evol} which show
evolutions of the coarse variables when the system parameters
(temperature $T$ and chemical potential $\mu$) are switched from 
 $k_BT = 7.5$, $\mu = -46.20$ to $k_BT = 7.0$, $\mu =-47.40$.
The small cluster variables are shown in Fig.~\ref{F:small_evol}
and are extremely fast - they approach the slow 
manifold on the timescale of less than 0.1 million steps.
The micelle size distribution is also relatively fast -- as
Figs.~\ref{F:micelle_evol}a and \ref{F:micelle_evol}b show, 
equilibration time of the first two moments of the micelle size distribution
is on the order of 10 million steps.
%
The slowest dynamics is that of the micelle number distribution 
-- it equilibrates on the order of 1000 million steps 
(see Figs.~\ref{F:micelle_evol}c and \ref{F:micelle_evol}d).
The observed timescales suggest that the coarse projective integration
should be performed for only two variables --  $\nu_m$ and $\sigma_m$ --
average and standard deviation of number of micelles.

\begin{figure}
\includegraphics[width=5in]{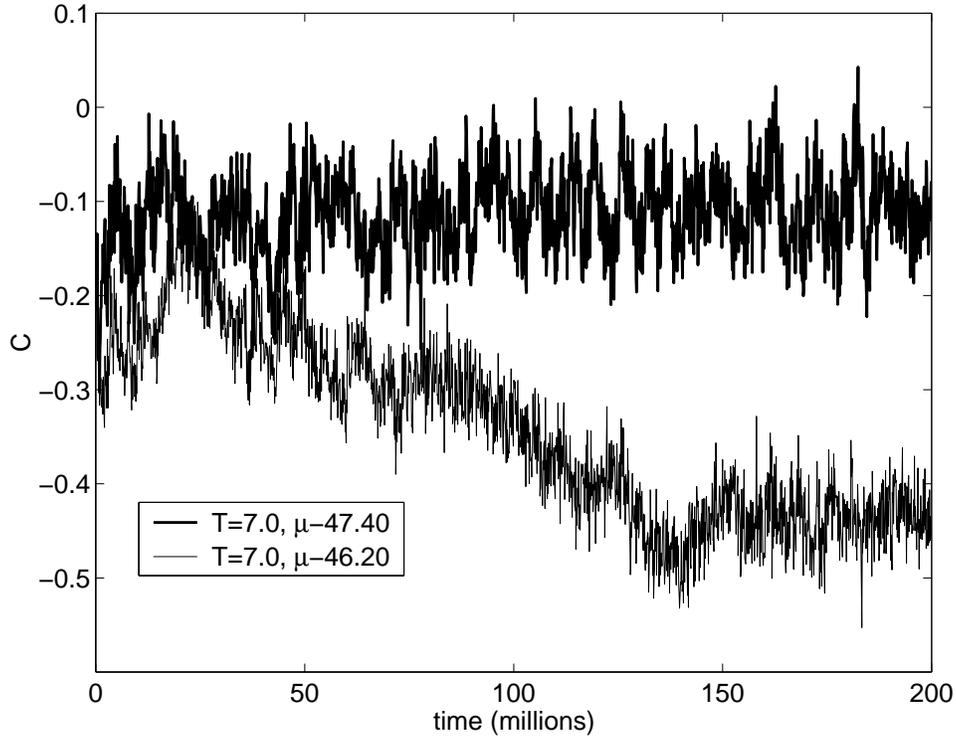}
\caption{Correlation coefficient between $\Nmicelle$ and the average 
         micelle size.
	 Shown are simulation results for two sets of parameter values:
	 low surfactant density, $k_BT = 7.0, \mu = -47.40$ (thick line) and  
         high surfactant density, $k_BT = 7.0, \mu = -46.20$ (thin line).     
         Both of these simulations have been initialized from a configuration at
	 $k_BT = 7.5, \mu = -46.20$. The reported correlation functions
	 are obtained from averaging of 500 MC realizations.
}
\label{F:Nmic_corr}
\end{figure}

The described choice of coarse variables assumes that there is no 
correlation between the number of micelles $\Nmicelle$ in the system
and the micelle size. 
In order to check this assumption, we computed the
correlation coefficient between $\Nmicelle$ and the average size
of a micelle at different values of temperature $T$ and the 
chemical potential $\mu$. 
Results of this 
calculation are shown in Figure~\ref{F:Nmic_corr}. 
For relatively dense systems, the correlation coefficient is non-vanishing 
and we might need to take it into account when working at those conditions.
However, for low surfactant
density (thick line in Fig.~\ref{F:Nmic_corr}), 
the correlation is negligible.
The coarse integration results below are reported for this low density.

\section{The Lifting Procedure}
\label{S:lifting}

In this section, we describe the lifting procedure, i.e. how realizations of
the detailed system are reconstructed from the 7 coarse variables.
Since there is no long-range interaction between the clusters
in the system and we consider sufficiently dilute systems, the lifting
procedure consists of the following two stages: first, we generate a sequence 
of sizes of clusters to be placed into the system.
Second, we place these clusters into the simulation box.
In order to improve efficiency of the second stage of lifting, 
we sort the list of cluster sizes generated in the first stage 
in descending order. 
This is done
because it is easier to place a larger cluster into an emptier system
and it does not introduce any bias since correlations
between different clusters in the system are negligible.
For each cluster size from the list, we randomly pick a cluster from a database
and place it into a random location in the simulation box. 
The only 
requirement in this procedure is that the clusters do not overlap or 
touch each other. 
In principle, this procedure can be generalized to the case
of denser systems or systems with long-range interactions, 
where the correlations between clusters become important
(see also Ref.~\onlinecite{MMPK02}, where pair probabilities between
adsorbate atoms must be appropriately initialized).

Let us now describe the generation of the cluster size list in 
more detail. 
This procedure is split into two parts. 
First, the
micelle sizes are generated: the 
number of micelles $\Nmicelle$ in the system is sampled from
the truncated Gaussian distribution Eq. (\ref{eq:Gauss}) 
with mean $\nu_m$ and
standard deviation $\sigma_m$.
Then, for 
$j = 1, \dots, \Nmicelle$, the $j$-th micelle size is sampled from
the Gaussian distribution with parameters $\nu$ and $\sigma$.

The number $\Nsmall$ of molecules contained in small clusters is sampled
from the Gaussian
distribution with parameters $\sigma_s$ and $\nu_s$.
Then the cluster sizes $N_j$, $j = 1, 2, \dots$ are sampled
from the Poisson distribution Eq. (\ref{eq:Poisson}) with 
parameter $\lambda$. 
We stop after we have generated $M$ clusters such that
\begin{equation}
\label{eq:Nsmall}
   \sum_{j=1}^{M-1} N_j \le \Nsmall < \sum_{j=1}^M N_j.
\end{equation}
We then keep all of these clusters if the sum on the right hand side of 
Eq (\ref{eq:Nsmall}) is closer to $\Nsmall$ than the sum on the
left hand side. 
Otherwise, we keep only the first $M-1$ clusters generated.
Although this procedure does not always produce
exactly $\Nsmall$ molecules in each realization, it does so on
average.
Moreover, as we have shown, the small cluster variables are 
slaved to the micellar variables. 
Therefore, small errors in
small clusters lifting will be quickly healed in the microscopic
simulation.
If necessary, the short ``healing" step can be implemented
in a constrained fashion (evolving the MC simulation with a hard
parabolic potential around the target values of the slow observables,
as in umbrella sampling
\cite{Potential}).

\section{Coarse Projective Integration}
\label{S:Euler}
In this section we describe coarse integration of the micellar system
using the coarse projective Euler method.
This method is briefly outlined in the introduction and
is discussed in more detail by
Gear and Kevrekidis \cite{GK03a}.

Other coarse integration methods are also proposed in 
Ref.~\onlinecite{GK03a,RMGK04} and their linear stability
analysis is performed.

Before we start the coarse integration, we perform the microscopic (GCMC) 
simulation and let all the fast modes equilibrate. 
Fast modes here refer to the small cluster variables and the
first two moments of the distribution of micelle sizes. 
As shown in section \ref{S:coarse_var}, these modes are much 
faster than the mean $\nu_m$ and the variance $\sigma_m$ of 
the number of micelles in the simulation box.
After the ``fast mode equilibration" is complete, we let the 
simulations run for a little longer in order to collect enough samples
of clusters for the database. 
We run this ``database production"
simulation for an additional 5 million steps and use 500 copies of the system.
The database is updated after each 0.1 million steps. 
This step size is
sufficiently large for the micelle structures to be significantly modified
and hence we populate the database with statistically different clusters. 
A more detailed analysis of the rate of change of a cluster 
structure will be presented in the companion paper \cite{KPK04b}. 

When the initial preparations are complete, we perform the coarse projection.
The lifting procedure was described in the previous section. 
After the lifting, we let the system ``heal" for $\theal = 0.2$
million steps. 
This time is more than sufficient to eliminate any
discrepancies in the (fast) small cluster variables
$\lambda$, $\nu_s$, and $\sigma_s$ that might be introduced during 
the lifting
-- see the earlier discussion of timescales. 
Since we pick the micelle
structures directly from the ``equilibrium" database, we expect that 
no healing is needed for the micelle size variables $\nu$ and 
$\sigma$.
As mentioned above, the initial ``healing" preparatory step can also be implemented
in a constrained fashion as in umbrella sampling
\cite{Potential}.

After the healing is complete, we perform the (unconstrained) simulation for 
additional $\tgen = 9.8$
million steps. 
We then fit a straight line through the computed 
$\nu_m(t)$ and $\sigma_m(t)$ and extrapolate these quantities for
the macroscopic step $\Delta t = 50$ million steps,
\begin{equation}
   \bx(t+\Delta t) = \bx(t) + \Delta t \bbf(\bx(t)).
\end{equation}
Here, $\bx(t) = (\nu_m(t), \sigma_m(t))$ are the slow coarse
variables and $\bbf(\bx(t))$ is their slope obtained from the
least squares fit to the results of the production run of length
$\tgen = 9.8$ million steps.
The fast coarse variables are set to the values from 
the last step of the previous microscopic simulation run.

\begin{figure}
\includegraphics[width=5in]{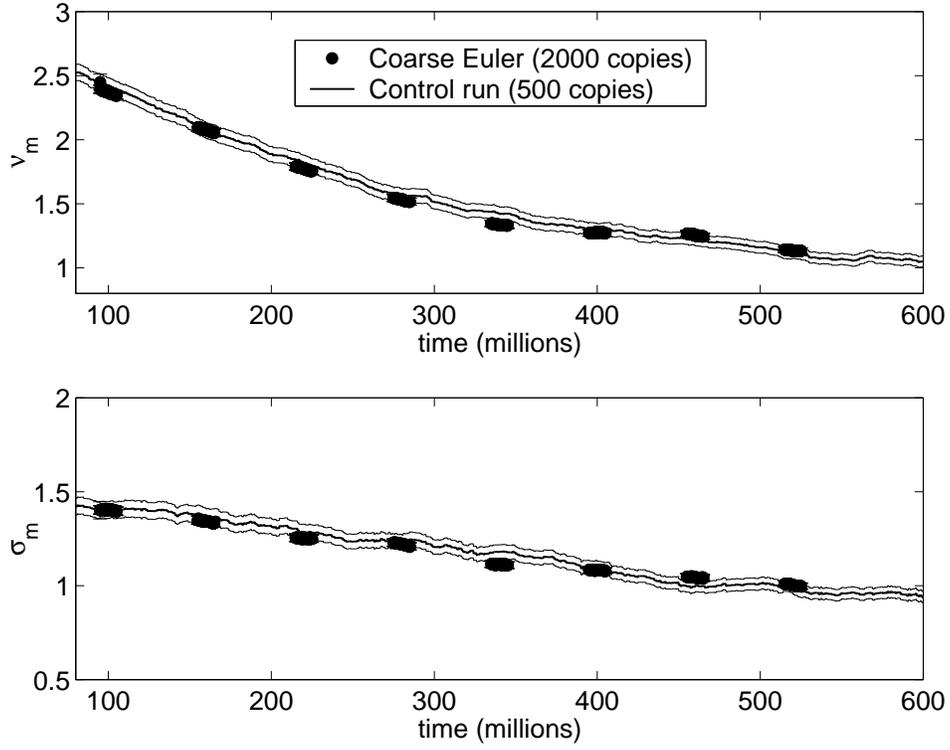}
\caption{Results of coarse Euler method: the solid lines
         correspond to 
         a control run (thin lines show the error bars), and the circles 
	 show the evolution
	 of projected coarse variables; $\theal=0.2$ million, $\tgen=9.8$ 
	 million, $\Delta t = 50$ million. The error bars for the coarse
	 integration are relatively small and are almost invisible in
	 the plot.}
\label{F:euler}
\end{figure}

Results of the coarse integration are shown in Fig.~\ref{F:euler}.
We observe good agreement -at the level of the macroscopic observables-
with the control full scale GCMC run, also shown in this figure.
These simulations are performed
for a system switching from $k_BT = 7.5$, 
$\mu = -46.20$ to $k_BT = 7.0$, $\mu = -47.40$.
We used different numbers of copies of the system in the 
coarse integration and the control run: in the coarse integration, 
we used 2000 copies and in the control run we used 500 copies.
The number of copies had to be increased for the coarse integration in
order to reduce statistical error in the extrapolations:
the projection step ``magnifies" the noise and one has to have 
precise data in order to obtain a reasonable accuracy during extrapolation.
Hence, there is a trade off between the size of the projection step and
number of copies of the system needed for an accurate extrapolation.

In our particular case, the savings of CPU time in the coarse integration
(as compared to the control run) are about 50\%.
The efficiency of the coarse Euler scheme is the subject of further investigation
(see Section~\ref{S:KMC}) and possible improvements to the method
are discussed in Section~\ref{S:Discussion}.


\section{Coarse Newton Method}
In addition to coarse projective integration, one can perform other system-level tasks,
such as fixed point location and stability analysis, using the information obtained
from the short-scale simulations. 
In this section, we describe an application of a Newton-like method to the micellar system. 
The approach is based on the observation that the
equilibrium configuration of the system corresponds to a solution of a 
nonlinear system of equations
\begin{equation}
   \bbf(\bx) = 0,
\end{equation}
where $F_i$ is the slope of the evolution of the $i$-th coarse variable 
$X_i(t)$.
The slope of $\bbf$ is estimated by fitting a straight line to 
results of the short-scale microscopic simulations.
Hence, one can use the Newton method (or some other method of solution
of nonlinear equations) in order to obtain the equilibrium configuration
of the system.

In a Newton algorithm for a deterministic problem one evaluates the
residual at the current guess, the Jacobian of the equations at the
current guess, and solves a linear set of equations to provide the
next guess, at which the procedure is repeated. 
The construction of variants of the Newton method appropriate for 
fixed point computations in noisy environment is the subject of
ongoing research (e.g. the stochastic approximation algorithms 
\cite{SA1,SA2}).

In this paper we try to estimate the deterministic component
of the noisy simulation, reducing the variance through 
averaging a number of copies, and perform Newton-Raphson
on the deterministic part.
The derivatives required in the Newton step will be estimated
using finite differences (i.e. by initializing {\it macroscopically
nearby} initial conditions, and observing the difference in the
evolution of their coarse variables).

One important practical twist in our implementation of the
Newton method for a noisy/stochastic system (compared
to deterministic noise-free system of equations)
is that our lifting procedure does not create microscopic
states {\it precisely} consistent with the macroscopic
observables; rather, it creates microscopic states
consistent with slightly {\it nearby} observables.
In the Newton context, the function $\bbf$ is computed not for 
the specified observable values $\bx$ of the coarse variables 
but for slightly  different values.
This happens  due to the stochastic nature of our lifting procedure,
which involves sampling of the
random variables according to a prescribed distribution.
Of course, if the sample size is infinitely large, then
the moments of the sampled realizations of the system will
coincide with the prescribed ones. However, the finite sample
size introduces some statistical errors in the generated moments.
It is also possible, {\it after} lifting and {\it before} the 
Newton procedure, to apply an additional preparatory step, in 
which the microscopic state is adjusted to correspond to
the macroscopic observables exactly (either through constrained
evolution or some sort of simulated annealing). 
Such reasonable modifications of the procedure are the subject
of current research; we are proceeding with our current lifting
operator.

To be more specific, consider the lifting procedure for micellar 
systems, which involves sampling of the number of
micelles $\Nmicelle$ from the first two moments of the 
truncated Gaussian distribution. I.e. in this case 
the coarse variables are
$\bx = (\nu_m, \sigma_m)$. 
We observe that for the values of $\nu_m$ and
$\sigma_m$ typical for our simulations, one
needs at least $10^4$ realizations in order to sample an accurate mean 
$\nu_m$ and 
standard deviation $\sigma_m$. 
This large value of realizations is 
impractical for the micellar simulations and we have used 2000 copies
in our implementation of the Newton method.

The implication of this statistical error with our current lifting is that it is 
necessary to modify the standard finite-difference calculations of 
the Jacobian. 
In what follows, we first describe the necessary changes 
for a one-dimensional function
and then generalize them to a system of arbitrary dimensionality.
Consider the standard forward difference estimation of a
derivative,
\begin{equation}
\label{eq:deriv_standard}
  \frac{dF(x)}{dx} = \frac{F(x+\Delta x)-F(x)}{\Delta x} + O(\Delta x)
\end{equation}
However, in the lifting, instead of $x$ and $x+\Delta x$, we
generate $L(x)$ and $L(x+\Delta x)$, where $L(x)$
denotes the value of a coarse variable which is generated during
the lifting step from the prescribed value $x$. 
We expect that, as the number of 
copies of the system approaches infinity, $L(x)$ approaches $x$ 
and therefore, for a finite but
large number of copies, $L(x)$ is not very different from $x$.
Hence, instead of $F(x)$ and $F(x+\Delta x)$, we compute 
$F(L(x))$ and $F(L(x+\Delta x))$ and thus the correct forward
difference estimate for the derivative is
\begin{equation}
\label{eq:deriv_correct}
  \frac{dF(x)}{dx} = \frac{F(L(x+\Delta x))-F(L(x))}{L(x+\Delta x)-L(x)} +
                     O(\Delta x).
\end{equation}
We found that this correction to the standard forward difference formula
(\ref{eq:deriv_standard}) is significant in 
the implementation of the coarse Newton method 
for the micellar system and that taking it into account
have improved the convergence of the method.

The correction to the forward-difference scheme described above
is the simplest estimate for the derivative in our noisy system.
Without more precise lifting, we do not expect standard higher order
schemes to yield significant improvement in the Newton convergence.
%
%
There is significant recent interest in development of higher order
difference schemes for derivative estimation in noisy systems. 
In particular, Gear \cite{Gear_Newton02} 
has proposed to use the least-squares fit and Drews \emph{et al.} 
\cite{Braatz03}  have developed 
a central difference scheme 
with the weights chosen
in order to reduce the variance of the derivative estimate.
Their estimate is still $O(\Delta x)$ but the variance of 
$F'(x)$ is reduced. The work on testing and developing such schemes
is currently in progress. 

In this paper, we use the corrected
forward-difference scheme Eq. (\ref{eq:deriv_correct}).
Let us briefly discuss the generalization of this formula to 
multiple dimensions
while keeping in mind the standard forward-difference scheme,
\begin{equation}
\label{eq:Jacoby}
   \frac{\partial F_i(\bx)}{\partial x_j} = 
   \frac{F_i(\bx + \Delta x_j\bej)-F_i(\bx)}{\Delta x_j} +
   O(\Vert\bx\Vert), \quad i,j = 1, \dots, N.
\end{equation}
Here, $N$ is the dimensionality of the (coarse) system, $\bej$ is a unit vector pointing in 
the $j$-th direction and
$\Delta x_j$ is an increment in $x_j$. 
Due to the uncertainties in 
the lifting procedure, we actually compute the function $\bbf(\bx)$ at
points $\bxzero \equiv L(\bx)$ and $\bxj \equiv L(\bx + \Delta x_j\bej)$.
From the Taylor expansion of the function $F_i(\bx)$,
\begin{equation}
   F_i(\bxj) = F_i(\bxzero) + \nabla F_i(\bxzero) \cdot (\bxj-\bxzero) + 
                O(\Vert(\bxj-\bxzero\Vert),
\end{equation}
it follows that 
\begin{equation}
\label{eq:deriv_multidim}
   \nabla F_i(\bxzero) = A^{-1} \bg,
\end{equation}
where $O(\Vert(\bxj-\bxzero\Vert)$ terms have been neglected and
$g_j \equiv F_i(\bxj)-F_i(\bxzero)$ and $A_{kj} \equiv x_{kj}-x_{0j}$,
$j,k = 1, \dots, N$
($x_{kj}$ denotes the $k$-th component of the vector $\bxj$).
This formula, Eq. (\ref{eq:deriv_multidim}), is similar to 
the forward difference simplex gradient described, e.g., in 
Ref.~\onlinecite{Kelley_optimization_book} in the context of
optimization of noisy functions.
It is possible to use this formula with less or more
than $N+1$ evaluations of function $\bbf(\bx)$. 
In the first case,
the solution of Eq. (\ref{eq:deriv_multidim}) will be given by
a pseudoinverse of the matrix $A$ and the computation will be 
somewhat similar to the simultaneous perturbation
stochastic approximation method \cite{Spall00}.

In the second case, the estimate of the Jacobian will be given
by the least squares fit \cite{Gear_Newton02}.

In the current work, we limit the estimation of the derivative 
to the direct use of Eq (\ref{eq:deriv_multidim}) and hence
we perform $(N+1)$ evaluations of the left-hand-side per iteration
of the Newton method.
We choose the increment $\Delta x_j$ of $x_j$ in the calculation of the
derivative to be an integer multiple of the statistical
error $x_{err}$ of $x_j$. 
%
This choice assures that the increments of different coarse
variables are not too small (i.e. not smaller than the
noise).
The stopping criterion for the iterations is
\begin{equation}
   F_i(x) \le F_{i,\,err}(x),\ i = 1, \dots, N,
\end{equation}
where $F_{i,\,err}$ is the statistical error estimate for the function
$F_i$.

We perform coarse Newton simulations for the two slowest 
coarse variables, $\bx(t) = (\nu_m,\sigma_m).$
The function $\bbf(\bx)$, i.e. the slope of $\bx$(t), is computed 
by fitting a straight line to result of an ``inner" simulation
of 2000 copies of the system. 
The inner simulation is 
performed for 10 million steps:
the first 0.2 million steps are used for healing and 
the linear fit is performed for the last 9.8 million of steps of the run.
The values of the fast coarse variables are initialized at the values
at the end at the end of the previous inner integrator run.

\begin{figure}
\includegraphics[width=5in]{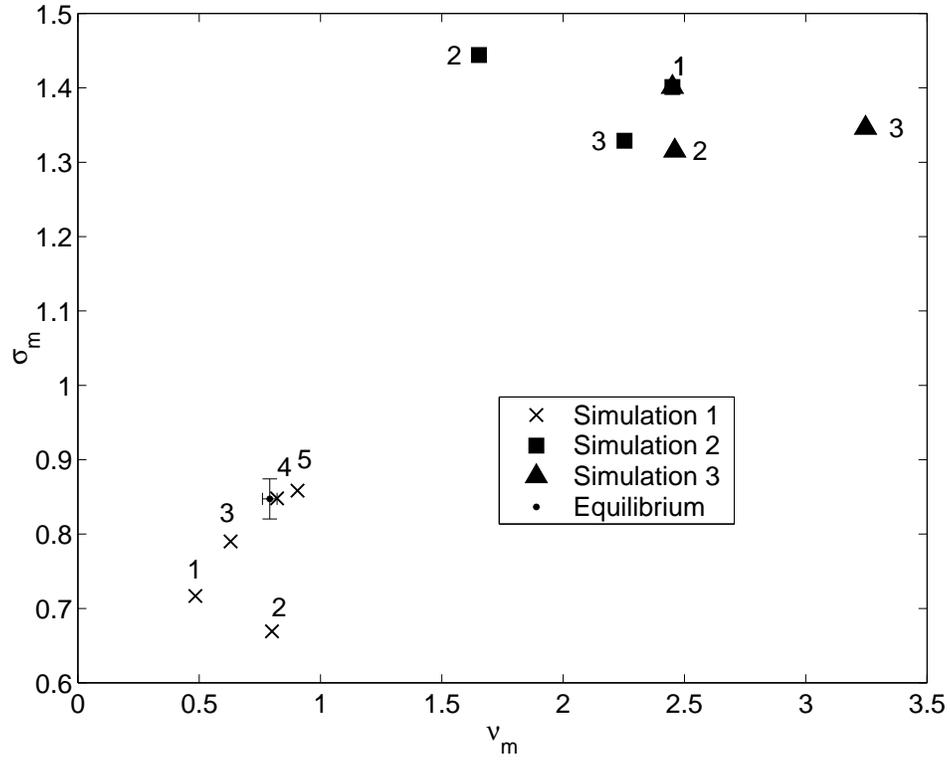}
\caption{Results of the coarse Newton method. The numbers on the plot 
indicate the iteration number. See text for details.}
\label{F:newton}
\end{figure}

Results of 3 calculations using the Newton method are shown in 
Fig.~\ref{F:newton}.
An estimate of the equilibrium solution obtained from an equilibrium
run with 500 copies of the system is shown by the circle.
The 3 coarse Newton simulations are as follows: 
\begin{enumerate}
\item Simulation 1 (crosses): the iterations started relatively close
      to the solution; in the derivative estimation, the increment for
      derivative calculation is $\Delta x = 5 x_{err}$.
\item Simulation 2 (squares): the iterations started at the same
      point at which the coarse Euler simulation started,
      see Fig.~\ref{F:euler}; $\Delta x = 5 x_{err}$.
\item Simulation 3 (triangles): the iterations started at the same
      point as those of the second run but now the derivative
      is computed with $\Delta x = x_{err}$.
\end{enumerate}

The first Simulation converges to a point located close to the estimation
of the equilibrium solution. 
The slight difference between the equilibrium
estimate and the coarse Newton method result is probably due to fact that 
500 copies (used in the control run)
of the system do not provide as good statistics as 2000 
copies (used in coarse Newton) do.

However, the situation is worse for the Simulations 2 and 3
started relatively far from the stationary point. 
These two
calculations were terminated after the third iteration because \
it did not appear that the iterations were converging to a stationary
solution. 
Moreover, 
these iterations are very sensitive to the choice of $\Delta x$: even
the direction of Newton step is altered by changing $\Delta x$ by a 
factor of 5. 
One reason for such a poor performance of the Newton method
far away from the stationary point is the high level of noise.
In section \ref{S:KMC} we show that the function $\bbf(\bx)$ is very 
noisy if one only uses 2000 copies of the system.
It is well known, from the simple deterministic context, that Newton
iteration is quite sensitive (in a problem-dependent way) to the choice
of initial conditions.
It is mostly in a continuation context (i.e. when we have the solution
at one parameter value, and want to find it at nearby parameter values)
that Newton is routinely used.
In the context of the present work, we only want to demonstrate that
such continuum-inspired algorithms can work when applied appropriately
and with sufficient variance reduction.
Many deterministic variants of the Newton (including matrix-free
Newton-Krylov methods) can be easily modified to work in a coarse
timestepper context; which of these will be the least sensitive to
noise, and what savings they can produce is the subject of current
research.
Once more, what we want to show here is that such continuum numerical
analysis methods are, indeed, applicable as ``wrappers" around the
type of atomistic micellar simulations we perform.
It is also important to notice that algorithms like Newton are
capable of converging to unstable stationary states (e.g. transition
states) and can be augmented to converge on {\it marginally stable}
states (at the onset of instabilities) \cite{MMK02,SGK03,HK03,SMK04}.

In addition to computing an equilibrium configuration of the system,
results of the Newton method can be used to obtain the timescale 
(in MC iteration time) of approach of the coarse variables
to this equilibrium. 
Eigenvalues of the coarse Jacobian near
a stationary point correspond to timescales of the coarse variables.
For the density variables, $\nu_m$ and $\sigma^2_m$, the eigenvalues
computed near the equilibrium solution are on the order of $10^{-8}$.
This is in contrast with the eigenvalue $5 \times 10^{-6}$
obtained in the companion paper
for the evolution of a single micelle near its equilibrium configuration.
These calculations confirm that there is
at least a two order of magnitude separation of times cales between
the coarse-grained dynamics of micelle density and the micelle size variables.

\section{Kinetic Monte Carlo model}
\label{S:KMC}

In this section, we show that the poor performance of the Newton
method is due to the high level of noise in the system.
The simplest way to estimate the noise level in the evaluations
of function $\bbf(\bx)$ is to compute values of this
function on a relatively fine mesh of values of $\bx$. 
Unfortunately, this task is formidable for our micellar system --
a calculation of the function $\bbf(\bx)$ at a single point takes 
about 28 hours on a single AMD Athlon processor and, even with 
parallelization, the calculation on a fine mesh would take very 
long time.

In order to compute $\bbf(\bx)$ on a mesh and to prepare the ground 
for future ``experiments" with various variance-reducing finite-difference
methods, we consider the following simple kinetic Monte Carlo (KMC) model.
We assume that there is a discrete set of possible states of the system and
we number these states $0, 1, \dots, N$.
In the case of a micellar system, the $i$-th state 
corresponds to the system with $i$ micelles in it.
We consider $\nrepeat$ copies of the system and store
the data in a form of a histogram 
$\bh = (h_0, h_1, \dots, h_N)$ so that the value of $h_i$
is the number of copies of the system which are in the $i$-th state and
$\sum_{i = 0}^N h_i = \nrepeat$.
We further assume that the birth-death process is described by a first-order
master equation,
\begin{equation}
\label{eq:master}
   \frac{d\bh}{dt} = \bk\bh,
\end{equation}
where $\bk$ is the matrix of transition rates and $k_{ij}$ is the rate of
transition from state $i$ to state $j$. 
It is shown in the companion paper that 
the birth-death of micelles can indeed be approximated by the first
order kinetic process.

Although the solution of equation (\ref{eq:master}) is straightforward,
in some cases this equation is not available in an explicit form.
This happens, in particular, when it is not easy to identify the states 
of the system or the transition paths between these states.
Examples of such systems include
diffusion in random media and birth and death of micelles in a system
with long-range interactions.
In these cases, one has to resort to some microscopic integration tool
such as MD or MC instead of solving the more macroscopic master equation.

In order to model the birth-death process and its interaction with the coarse
integration tools, we use stochastic (kinetic Monte Carlo, KMC) simulation 
\cite{Kang89,Chakraborty93,SnurrTST} 
to simulate the equation (\ref{eq:master}). 
In our simulations, we choose KMC parameters which model micelle
birth/death process. 
The transition rates (i.e. rates of micelle birth and death)
are chosen so that they satisfy the detailed balance condition
consistent with the equilibrium distribution $\Peq(i)$,
i.e.
\begin{equation}
\label{eq:detailed_balance}
   \frac{\Peq(i-1)}{\Peq(i)} = \frac{k_{i,i-1}}{k_{i-1,i}}.
\end{equation}
The equilibrium distribution is obtained from the full scale 
MC simulation.

\begin{figure}
\includegraphics[width=5in]{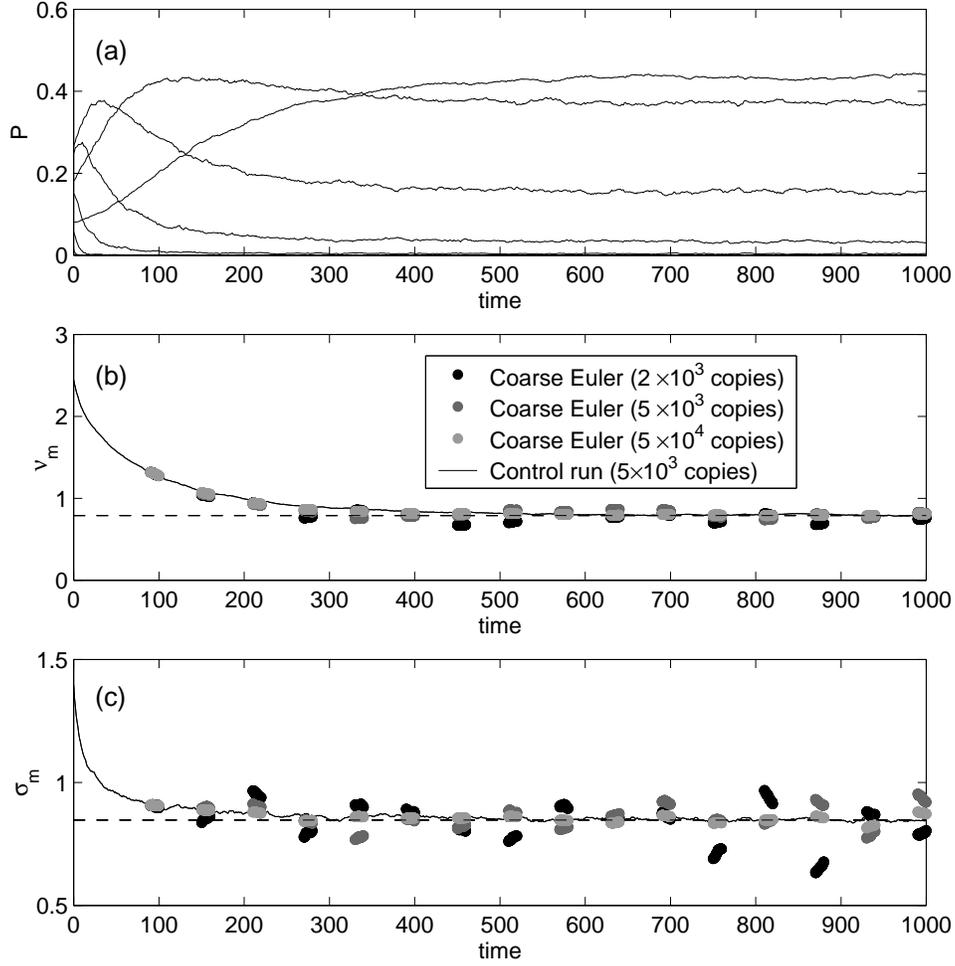}
\caption{Results of KMC simulations with $\nrepeat=5000$ copies 
         of the system (solid lines): 
	 (a) evolution of probability distribution
	 $P(i)$, each line corresponds to a different state $i$,
	 $i = 0, \dots, 7$; (b) evolution of the mean $\nu_m$, and
	 (c) evolution of the standard deviation $\sigma_m$.
	 Time here is measured in millions of GCMC steps.
	 Circles in plots (b) and (c) show results of the coarse
	 projective Euler method for $\nrepeat = 2\times 10^3$
	 copies (black circles), $\nrepeat = 5\times 10^3$ copies
	 (dark gray circles), and $\nrepeat = 5\times 10^4$ copies
	 (light gray circles).
	 }
\label{F:euler_kmc}
\end{figure}

The condition (\ref{eq:detailed_balance})
guarantees convergence of the system to the 
equilibrium distribution.
The rates of the system are chosen so that one unit of time of the 
KMC simulation corresponds to one million of GCMC steps.
An example of a KMC simulation with thus chosen transition rates
is shown in Fig.~\ref{F:euler_kmc}. 
Fig.~\ref{F:euler_kmc}a shows evolution
of the probabilities $P(i) = h(i)/\nrepeat$ and it is seen that these
probabilities do approach an equilibrium.
Evolution of the two moments (mean $\nu_m$ and standard deviation
$\sigma_m$) of this distribution is shown in Fig.~\ref{F:euler_kmc}b and 
\ref{F:euler_kmc}c.
Comparison of Fig.~\ref{F:euler} and Fig.~\ref{F:euler_kmc} shows that
the KMC model reproduces essential features of the evolution and
hence its results can be used to analyse the accuracy
coarse integration technique.
Thus, in effect, in this model we have completely eliminated the 
fast dynamics and we focus on the rate-limiting slow dynamics --
birth and death of micelles. 
We emphasize that ``time" and ``dynamics"
here (as everywhere else in the paper) refer to the artificial
GCMC dynamics.

Probably the most important feature of equation-free computation
is that the algorithms are in effect ``wrappers" which, through
lifting and restriction subroutines, can be combined with {\it any}
microscopic/atomistic simulator.
In the same way we wrapped coarse projective integration and coarse
Newton around the GCMC with our 7 coarse observables, we will now
wrap it around the model kMC simulation (templated on the GCMC).
We perform coarse projective Euler integration for the KMC model with 3 different
numbers $\nrepeat$ of copies of the system.
The observables here are the average and the standard deviation
of the average $\nu_m$ and the standard deviation $\sigma_m$ 
of the number of micelles, just as in the coarse integration
of the full GCMC model.
Results of the coarse kMC integration are shown 
in Figs.~\ref{F:euler_kmc}b and c by circles. 
The timestep
of the Euler method is $\Delta t = 50$ and, before each projective
step, the inner integration is performed for 10 units of time.
The simulations with larger number of system copies ($5 \times 10^3$
and $5\times 10^4$) converge to the equilibrium solution. 
However, the simulation with $2 \times 10^3$ system copies 
exhibits strong oscillations around the stationary solution.


As evident from Figs.~\ref{F:euler_kmc}b and c, the accuracy
of the coarse Euler method significantly depends on the number 
$\nrepeat$ of the system
copies used in the microscopic simulation.
In addition to $\nrepeat$, the 
parameters of the coarse Euler method that can significantly
affect the accuracy are the 
the length $L$ of the microscopic simulation and
a length $\alpha$ of the projection interval.
For convenience, here we normalize $\alpha$ so that $\alpha = 1$ 
corresponds
to projection over the internal microscopic simulator step $\Delta t$.

In order to investigate the accuracy of the coarse Euler 
projective step, we obtain evolution of coarse variables 
from the KMC simulation and perform the
extrapolation of the coarse variables to various 
(normalized) timesteps $\alpha$.
We perform this projection for various values of 
realizations of the system $\nrepeat$ and for different lengths of the
fitting interval $L$. For each set of parameters $(\alpha, \nrepeat, L)$,
we compute $10^4$ realization of the ``average trajectory", i.e. 
a trajectory which one obtains using $\nrepeat$ copies of the KMC 
system. 
These average trajectories are still noisy and hence the
projections obtained from different realizations of these trajectories
will have some scatter. 
We measure this scatter by computing the
variance of $10^4$ realizations of such projections.
This variance, which we denote $\sigma^2(\alpha,\nrepeat,L)$,
measures the accuracy of the Euler step.

\begin{figure}
\includegraphics[width=5in]{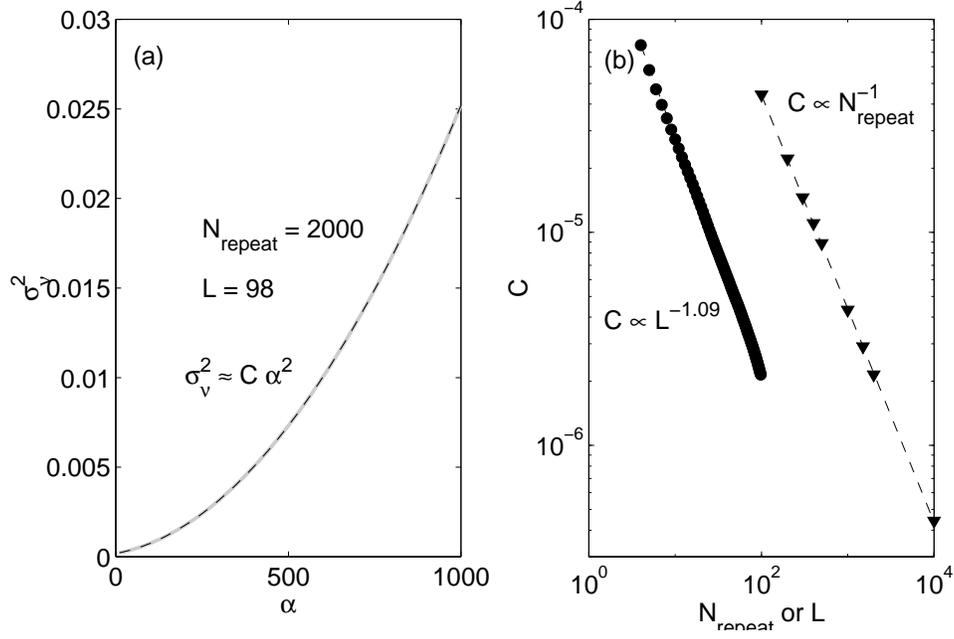}
\caption{(a) Estimated variance $\sigma^2_\nu(\alpha)$ of projection
         for $\nrepeat = 2000$, $L = 98$ as a function of the
	 normalized extrapolation time $\alpha$;
	 the thick gray line shows the KMC results and the thin
	 black dashed line shows the quadratic fit;
	 (b) triangles: scaling of the coefficient $C$ 
	 with respect to $\nrepeat$, $L=98$ is fixed;
	 circles: scaling of $C$ with respect to $L$;
	 $\nrepeat = 2000$ is fixed. The dashed lines 
	 show the logarithmic fit.}
\label{F:proj_kmc}
\end{figure}

The value of $\nrepeat$ was varied from $10^2$ to $10^4$, $\alpha$
was varied from $10$ to $10^3$, and $L$ was varied from 4 to 98.
Initial conditions for the KMC simulation were taken to coincide
with those of the simulations shown in Fig.~\ref{F:euler_kmc}.
Here, we report results of the projection of $\nu_m$.
The conclusions for the other coarse variable, $\sigma_m$, are 
similar.
Dependence of the variance of $\nu_m$, $\sigma^2_\nu(\alpha,\nrepeat,L)$, 
on the normalized projection time $\alpha$
is shown in Fig.~\ref{F:proj_kmc}a for $\nrepeat = 2000$ and $L = 98$. 
The quadratic dependence of $\sigma^2_\nu$ on $\alpha$ seen in 
Fig.~\ref{F:proj_kmc}a is also
observed for all values of $\nrepeat$ and $L$ considered in our 
numerical experiments.
We conclude therefore that
\begin{equation}
   \sigma^2(\alpha,\nrepeat,L) = C(\nrepeat,L) \alpha^2,
\end{equation}
where $C(\nrepeat,L)$ is the proportionality coefficient. We plot 
$C(\nrepeat,L)$ for
fixed $L = 98$ and various $\nrepeat$ as well as for fixed $\nrepeat=2000$ and
various $L$ in Fig.~\ref{F:proj_kmc}b and observe that, to a good
accuracy, $C$ is inversely proportional to both $\nrepeat$ and $L$.
We therefore conclude that
\begin{equation}
\label{eq:efficiency}
   \sigma_\nu^2(\alpha,\nrepeat,L) \approx \frac{\alpha^2}{L\nrepeat}.
\end{equation}
Similar result is obtained for the accuracy of the projection of 
$\sigma_m$.
The significance of the scaling Eq (\ref{eq:efficiency}) 
is that there is a trade-off between
the projection interval length $\alpha$ and the number of copies
$\nrepeat$ and the interval length $L$ used 
in the microscopic simulations. In other words, 
if we increase the projection length $\alpha$, then we also need
to increase the product $NL$ by the same factor
in order to keep variance of the projected data at the same level as the variance
of the original data. 

The efficiency of coarse projective integration, the choice of filtering,
variance reduction \cite{Oettinger} and extrapolation techniques it 
can be combined with, is the subject of current study; our group
as well as other groups are comparing the efficiency of direct simulation
to that of coarse integration methods.



{\bf Coarse Newton method for the kMC Model.}
We perform the coarse Newton computations for 
$\nrepeat = 2\times 10^3$ (the number of copies used in the coarse 
Newton method for micelles) and $5\times10^4$
(some large number that is unrealistic for the GCMC simulations).
We perform the simulations with forward and central differences.
In the case of the central differences, the Jacobian is obtained
from the solution of the Eq. (\ref{eq:deriv_multidim})
in the least squares sense. 
The increment $\Delta x$ for calculation of the Jacobian is chosen
as follows: for simulations with $\nrepeat = 2\times 10^3$,
we set $\Delta x = 5x_{err}$, where $x_{err}$ is the 
error estimate of a coarse variable $x$. 
This choice is similar to that used in the implementation of the 
Newton method for GCMC simulations. 
However, such a choice is not practical for simulations with
$\nrepeat = 5\times 10^4$, since in this case, the noise level
is relatively low which leads to  
very small values of $\Delta x = 5 x_{err}$.
Hence, in the simulations with $\nrepeat = 5\times 10^4$, 
we use increment $\Delta x = 0.05$.
This value roughly corresponds to the 
increments used in the simulations with $\nrepeat = 2\times 10^3$.

\begin{figure}
\includegraphics[width=5in]{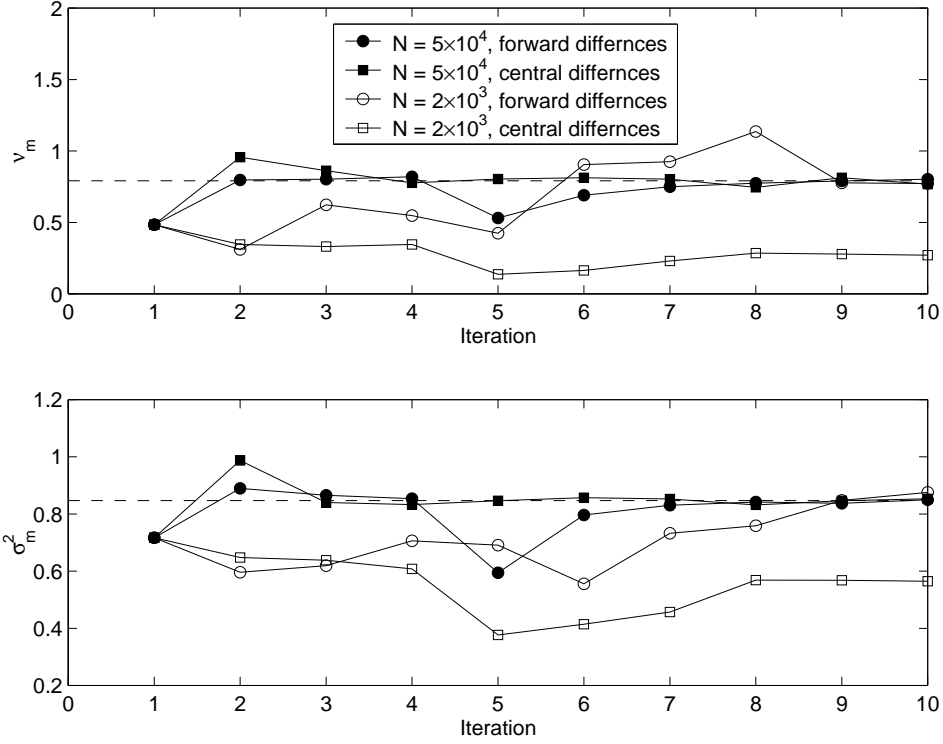}
\caption{Iterations of the Newton method for the KMC model. 
         Open symbols
         show simulations with $\nrepeat = 2\times10^3$ and closed
	 symbols show simulations with $\nrepeat = 5\times10^4$.
	 Circles show iterations of the Newton method with the forward
	 difference estimate of the Jacobian and squares show the
	 iterations with the central difference estimate of the
	 Jacobian.
	 Dashed lines show the estimate of the equilibrium.
	 Solid lines are shown to guide the eye.}
\label{F:newton_kmc1}
\end{figure}

\begin{figure}
\includegraphics[width=5in]{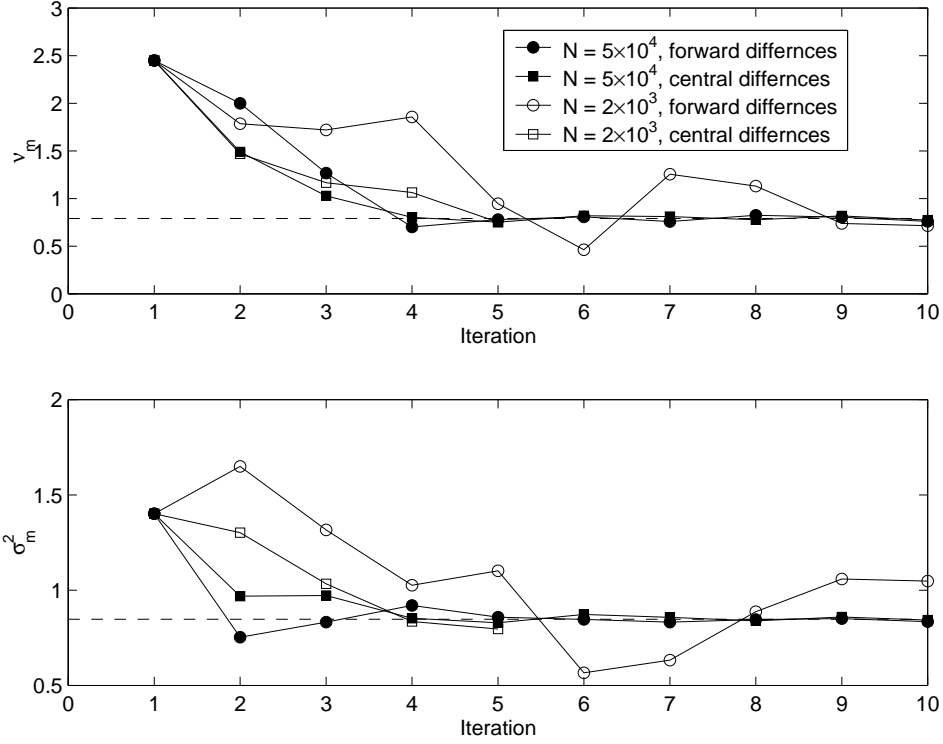}
\caption{Iterations of the Newton method for the KMC model.
         The same simulations as shown in Fig.~\ref{F:newton_kmc1}
	 except here the initial point is chosen further away
	 from the equilibrium solution.}
\label{F:newton_kmc2}
\end{figure}

Iterations of the coarse Newton method are shown in Figs.~\ref{F:newton_kmc1}
and \ref{F:newton_kmc2}. 
The simulations shown in Fig.~\ref{F:newton_kmc1},
are started from a point located relatively close to the equilibrium and
the simulations shown in Fig.~\ref{F:newton_kmc2} are started from a point
located further away from the equilibrium.
The iterations with $\nrepeat = 5\times 10^4$
converge relatively fast to the equilibrium and the convergence
is significantly faster if one uses the central difference algorithm.
However, for $\nrepeat = 2000$ (the number of copies which is realistic
in our GCMC simulations of the micellar system), the rate of convergence is
slower and, moreover, the central differences are not guaranteed to provide 
an improvement of convergence. 
In fact, in the simulations shown in Fig.~\ref{F:newton_kmc1}, 
the central
differences perform significantly worse than the forward differences.
Reliable estimation for the (coarse) derivative of
a noisy function is a vital element in the bridging of
microscopic similations with continuum-type numerical algorithms
based on (macroscopic) smoothness and Taylor series.

\begin{figure}
\includegraphics[width=5in]{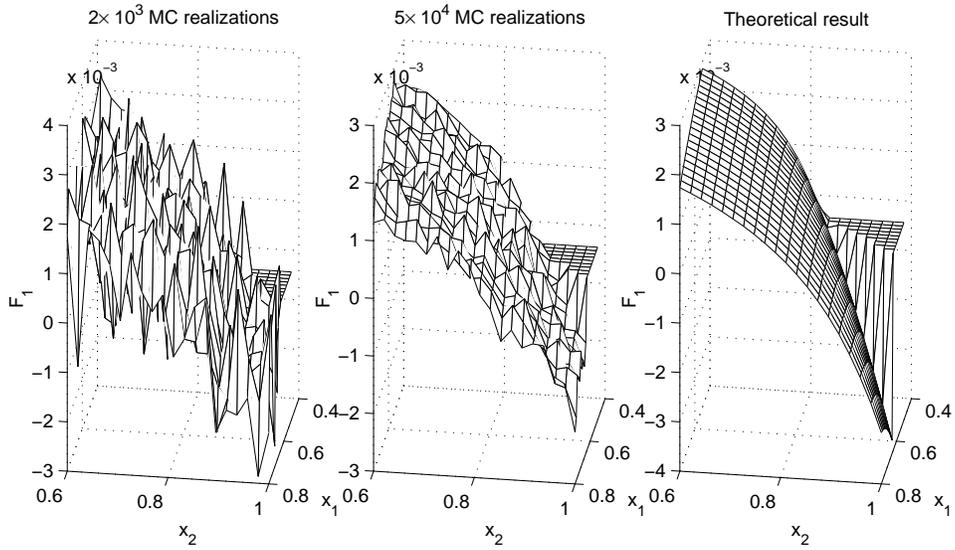}
\caption{$F_1$ obtained from KMC simulations and the 
         theoretical prediction}
\label{F:kmc_grid}
\end{figure}

Finally, in order to estimate the level of noise,
we compute one of the nonlinear functions ($F_1$ = slope of $\nu_m$)
for the Newton method and plot it in Fig.~\ref{F:kmc_grid} for
$\nrepeat = 2\times 10^3$ , $\nrepeat = 5\times 10^4$  and an analytic result obtained
from the solution of the master equation. 
We observe that the right-hand side is very noisy even for 
$\nrepeat = 5\times 10^4$ and is extremely noisy for $\nrepeat = 2\times 10^3$.
This explains such a poor performance of the coarse Newton method 
for $\nrepeat = 2\times 10^3$ and
suggest that a significant effort should be directed towards
developing variance-reducing schemes for the Newton method
\cite{Oettinger1,Spall00}.


\section{Discussion}
\label{S:Discussion}

We have demonstrated the application of coarse-grained, equation-free 
computational techniques to the GCMC simulations of the micellar system
(and to their kMC caricature). 
Of our 7 coarse variables, 5 ``fast" ones (small 
cluster parameters $\lambda$, $\nu_s$, and $\sigma_s$ and the micelle
size variables $\nu$ and $\sigma$) are observed to be slaved to two slow
(master) variables: average $\nu_m$ and the standard deviation $\sigma_m$
of the number of micelles in the system.

Although the current coarse computation results do not show significant
improvement of the efficiency of the coarse integration as compared to the
full-scale MC simulations, several possible improvements of the coarse 
integration methods are currently a subject of active research. 

Although in our micellar system, we have already taken advantage of the 
separation of time scales between the 5 fast coarse variables and the 2 slow
coarse variables, the dynamics of the 2 slow coarse variables ($\nu_m$
and $\sigma_m$) has not been fully explored. 
In particular, it is possible to improve 
the efficiency of the coarse computation by a different choice of the
slow coarse variables that describe the distribution $P(\Nmicelle)$ 
of the number $\Nmicelle$ of micelles in a simulation box.
Recall that in the current work, these variables are chosen to be 
the mean $\nu_m$ and the standard deviation $\sigma_m$ of $\Nmicelle$
and $\Nmicelle$ is assumed to have a truncated
Gaussian distribution. 
A more detailed analysis of the equation 
(\ref{eq:master}), which models evolution 
$P(\Nmicelle)$, shows a separation
of timescales which is absent in the dynamics of $\nu_m$ and $\sigma_m$. 
This may suggest different  coarse variables to represent $P(\Nmicelle)$,
that may better reflect the timescale separation in the evolution of
$P(\Nmicelle)$.

Another possibility for improvement of the accuracy of the coarse 
Euler method lies in the improvement of the projective step. 
In the current paper, we have used a linear least
squares fit to extrapolate the values of the coarse variables.
We assumed here that
the evolution of the coarse variables can be described by a deterministic 
equation. 
These deterministic coarse variables are ensemble 
averages of stochastic variables (such as number of micelles 
in a simulation box, considered in the current papers). 
It may be advantageous to take the stochasticity of this dynamics
into account in the projective integration step;
instead of fitting a deterministic model one may attempt to
fit a {\it stochastic} model for the evolution of the
observables, as is done in the companion paper \cite{KPK04b}.
Better extrapolation schemes (e.g. based on maximum likelihood
estimation and templated on Adams-Bashforth methods \cite{RMGK04})
which would incorporate filtering and take into account the correlations 
between values of the coarse variables at consecutive time steps,
may lead to more efficient algorithms.
Comparable statements also apply to the coarse Newton method. 

The point of this paper has been the illustration of a possible
bridge between traditional, continuum numerical methods and 
modern atomistic/microscopic simulations (here, GCMC simulations
of micelle formation)
Given the appropriate coarse observables, the detailed (here GCMC)
simulator is initialized {\it conditioned on} the observables
and then evolved for a (macroscopically) short time. 
The computational data are used to fit a local macroscopic evolution
equation which we assume exists.
In this paper we fit only the deterministic component of this equation;
more generally one can try to fit a local stochastic differential 
equation.
The local model is then used to design appropriate initial conditions
for new, subsequent computational experiments with the detailed 
simulator; the protocol for this ``design of computational experiments"
is provided by continuum numerical algorithms, such as initial value
problem solvers (e.g. forward Euler), fixed point solvers (e.g. Newton
Raphson) eigensolvers etc.
Smoothness (Taylor series) at the level of the macroscopic observables is
the underlying point of these methods; and what makes them possible is the
ability to initialize the microscopic simulator essentially ``at will",
consistent with macroscopic observables.

It is appropriate to close this paper with a short ``advertisement" for the
companion one \cite{KPK04b}.
In this paper we assumed an underlying smooth deterministic model for the
expected behavior of the macroscopic observables.
For certain stochastic systems (exemplified by a particle in a double
well under the effect of noise), the long term behavior of the system
statistics (approach to a final equilibrium density) 
may well be modeled by continuum equations.
One has to collect simulation data for a single particle over long
times, before one can observe the rate of density evolution.
For such systems, it may be more appropriate to fit the short term dynamics
in terms of a stochastic differential equation (e.g. a Langevin-type equation).
The long characteristic times for the equilibration of our simulation
are indeed governed by rare events (the formation and destruction of micelles).
In the companion paper we will show how to use similar ``coarse computation"
methods to design experiments based on an {\it effective stochastic evolution equation}
for the macroscopic observables, rather than the {\it effective deterministic evolution
equation} we used here.

{\bf Acknowledgements}. This work was partially supported by AFOSR and
an NSF/ITR grant. It is a pleasure to acknowledge discussions with 
Prof. C. W. Gear and Dr. G. Hummer.


\begin{thebibliography}{36}
\expandafter\ifx\csname natexlab\endcsname\relax\def\natexlab#1{#1}\fi
\expandafter\ifx\csname bibnamefont\endcsname\relax
  \def\bibnamefont#1{#1}\fi
\expandafter\ifx\csname bibfnamefont\endcsname\relax
  \def\bibfnamefont#1{#1}\fi
\expandafter\ifx\csname citenamefont\endcsname\relax
  \def\citenamefont#1{#1}\fi
\expandafter\ifx\csname url\endcsname\relax
  \def\url#1{\texttt{#1}}\fi
\expandafter\ifx\csname urlprefix\endcsname\relax\def\urlprefix{URL }\fi
\providecommand{\bibinfo}[2]{#2}
\providecommand{\eprint}[2][]{\url{#2}}

\bibitem[{\citenamefont{Theodoropoulos
  et~al.}(2000)\citenamefont{Theodoropoulos, Qian, and
  Kevrekidis}}]{Theodoropoulos00}
\bibinfo{author}{\bibfnamefont{C.}~\bibnamefont{Theodoropoulos}},
  \bibinfo{author}{\bibfnamefont{Y.-H.} \bibnamefont{Qian}}, \bibnamefont{and}
  \bibinfo{author}{\bibfnamefont{I.~G.} \bibnamefont{Kevrekidis}},
  \bibinfo{journal}{Proc. Natl. Acad. Sci.} \textbf{\bibinfo{volume}{97}},
  \bibinfo{pages}{9840} (\bibinfo{year}{2000}).

\bibitem[{\citenamefont{Siettos et~al.}(2003)\citenamefont{Siettos, Graham, and
  Kevrekidis}}]{SGK03}
\bibinfo{author}{\bibfnamefont{C.~I.} \bibnamefont{Siettos}},
  \bibinfo{author}{\bibfnamefont{M.~D.} \bibnamefont{Graham}},
  \bibnamefont{and} \bibinfo{author}{\bibfnamefont{I.~G.}
  \bibnamefont{Kevrekidis}}, \bibinfo{journal}{J. Chem. Phys.}
  \textbf{\bibinfo{volume}{118}}, \bibinfo{pages}{10149}
  (\bibinfo{year}{2003}), \bibinfo{note}{can be obtained as cond-mat/0211455 at
  arXiv.org}.

\bibitem[{\citenamefont{Hummer and Kevrekidis}(2003)}]{HK03}
\bibinfo{author}{\bibfnamefont{G.}~\bibnamefont{Hummer}} \bibnamefont{and}
  \bibinfo{author}{\bibfnamefont{I.~G.} \bibnamefont{Kevrekidis}},
  \bibinfo{journal}{J. Chem. Phys.} \textbf{\bibinfo{volume}{118}},
  \bibinfo{pages}{10762} (\bibinfo{year}{2003}), \bibinfo{note}{can be obtained
  as physics/0212108 at arXiv.org}.

\bibitem[{\citenamefont{J.~Cisternas and Kevrekidis}(2003)}]{CGLK04}
\bibinfo{author}{\bibfnamefont{S.~L.} \bibnamefont{J.~Cisternas},
  \bibfnamefont{C.~W.~Gear}} \bibnamefont{and}
  \bibinfo{author}{\bibfnamefont{I.~G.} \bibnamefont{Kevrekidis}},
  \bibinfo{journal}{submitted to Proc. Roy. Soc. London}
  (\bibinfo{year}{2003}), \bibinfo{note}{can be found as nlin.AO/0310011 at
  arXiv.org}.

\bibitem[{\citenamefont{Makeev et~al.}(2002{\natexlab{a}})\citenamefont{Makeev,
  Maroudas, and Kevrekidis}}]{MMK02}
\bibinfo{author}{\bibfnamefont{A.~G.} \bibnamefont{Makeev}},
  \bibinfo{author}{\bibfnamefont{D.}~\bibnamefont{Maroudas}}, \bibnamefont{and}
  \bibinfo{author}{\bibfnamefont{I.~G.} \bibnamefont{Kevrekidis}},
  \bibinfo{journal}{J. Chem. Phys.} \textbf{\bibinfo{volume}{116}},
  \bibinfo{pages}{10083} (\bibinfo{year}{2002}{\natexlab{a}}).

\bibitem[{\citenamefont{Makeev et~al.}(2002{\natexlab{b}})\citenamefont{Makeev,
  Maroudas, Panagiotopoulos, and Kevrekidis}}]{MMPK02}
\bibinfo{author}{\bibfnamefont{A.~G.} \bibnamefont{Makeev}},
  \bibinfo{author}{\bibfnamefont{D.}~\bibnamefont{Maroudas}},
  \bibinfo{author}{\bibfnamefont{A.~Z.} \bibnamefont{Panagiotopoulos}},
  \bibnamefont{and} \bibinfo{author}{\bibfnamefont{I.~G.}
  \bibnamefont{Kevrekidis}}, \bibinfo{journal}{J. Chem. Phys.}
  \textbf{\bibinfo{volume}{117}}, \bibinfo{pages}{8229}
  (\bibinfo{year}{2002}{\natexlab{b}}).

\bibitem[{\citenamefont{Kevrekidis et~al.}(2003)\citenamefont{Kevrekidis, Gear,
  Hyman, Kevrekidis, Runborg, and Theodoropoulos}}]{Manifesto}
\bibinfo{author}{\bibfnamefont{I.~G.} \bibnamefont{Kevrekidis}},
  \bibinfo{author}{\bibfnamefont{C.~W.} \bibnamefont{Gear}},
  \bibinfo{author}{\bibfnamefont{J.~M.} \bibnamefont{Hyman}},
  \bibinfo{author}{\bibfnamefont{P.~G.} \bibnamefont{Kevrekidis}},
  \bibinfo{author}{\bibfnamefont{O.}~\bibnamefont{Runborg}}, \bibnamefont{and}
  \bibinfo{author}{\bibfnamefont{K.}~\bibnamefont{Theodoropoulos}},
  \bibinfo{journal}{Comm. Math. Sciences} \textbf{\bibinfo{volume}{1}},
  \bibinfo{pages}{715} (\bibinfo{year}{2003}), \bibinfo{note}{original version
  can be obtained as physics/0209043 at arXiv.org}.

\bibitem[{\citenamefont{Kevrekidis et~al.}(2004)\citenamefont{Kevrekidis, Gear,
  and Hummer}}]{ShortManifesto}
\bibinfo{author}{\bibfnamefont{I.~G.} \bibnamefont{Kevrekidis}},
  \bibinfo{author}{\bibfnamefont{C.~W.} \bibnamefont{Gear}}, \bibnamefont{and}
  \bibinfo{author}{\bibfnamefont{G.}~\bibnamefont{Hummer}},
  \bibinfo{journal}{AIChE Journal} \textbf{\bibinfo{volume}{50}},
  \bibinfo{pages}{1346} (\bibinfo{year}{2004}).

\bibitem[{\citenamefont{Gear and Kevrekidis}(2003)}]{GK03a}
\bibinfo{author}{\bibfnamefont{C.~W.} \bibnamefont{Gear}} \bibnamefont{and}
  \bibinfo{author}{\bibfnamefont{I.~G.} \bibnamefont{Kevrekidis}},
  \bibinfo{journal}{SIAM J. Sci. Comput.} \textbf{\bibinfo{volume}{24}},
  \bibinfo{pages}{1091} (\bibinfo{year}{2003}), \bibinfo{note}{also NEC
  Technical Report NECI-TR 2001-029, can be obtained as
  http://www.neci.nj.nec.com/homepages/cwg/projective.pdf}.

\bibitem[{\citenamefont{Israelachvili}(1989)}]{Israelachvili_book}
\bibinfo{author}{\bibfnamefont{J.}~\bibnamefont{Israelachvili}},
  \emph{\bibinfo{title}{Intermolecular and Surface Forces}}
  (\bibinfo{publisher}{Wiley}, \bibinfo{address}{New York},
  \bibinfo{year}{1989}), \bibinfo{edition}{2nd} ed.

\bibitem[{\citenamefont{Larson et~al.}(1985)\citenamefont{Larson, Scriven, and
  Davis}}]{LarsonMC85}
\bibinfo{author}{\bibfnamefont{R.~G.} \bibnamefont{Larson}},
  \bibinfo{author}{\bibfnamefont{L.~E.} \bibnamefont{Scriven}},
  \bibnamefont{and} \bibinfo{author}{\bibfnamefont{H.~T.} \bibnamefont{Davis}},
  \bibinfo{journal}{J. Chem. Phys.} \textbf{\bibinfo{volume}{83}},
  \bibinfo{pages}{2411} (\bibinfo{year}{1985}).

\bibitem[{\citenamefont{Larson}(1996)}]{LarsonMC96}
\bibinfo{author}{\bibfnamefont{R.~G.} \bibnamefont{Larson}},
  \bibinfo{journal}{Journal de physique II} \textbf{\bibinfo{volume}{6}},
  \bibinfo{pages}{1441} (\bibinfo{year}{1996}).

\bibitem[{\citenamefont{Floriano et~al.}(1999)\citenamefont{Floriano,
  Caponetti, and Panagiotopoulos}}]{AZPmicelles99}
\bibinfo{author}{\bibfnamefont{M.~A.} \bibnamefont{Floriano}},
  \bibinfo{author}{\bibfnamefont{E.}~\bibnamefont{Caponetti}},
  \bibnamefont{and} \bibinfo{author}{\bibfnamefont{A.~Z.}
  \bibnamefont{Panagiotopoulos}}, \bibinfo{journal}{Langmuir}
  \textbf{\bibinfo{volume}{15}}, \bibinfo{pages}{3143} (\bibinfo{year}{1999}).

\bibitem[{\citenamefont{Panagiotopoulos
  et~al.}(2002)\citenamefont{Panagiotopoulos, Floriano, and
  Kumar}}]{AZPmicelles02a}
\bibinfo{author}{\bibfnamefont{A.~Z.} \bibnamefont{Panagiotopoulos}},
  \bibinfo{author}{\bibfnamefont{M.~A.} \bibnamefont{Floriano}},
  \bibnamefont{and} \bibinfo{author}{\bibfnamefont{S.~K.} \bibnamefont{Kumar}},
  \bibinfo{journal}{Langmuir} \textbf{\bibinfo{volume}{18}},
  \bibinfo{pages}{2940} (\bibinfo{year}{2002}).

\bibitem[{\citenamefont{Ferrenberg and Swendsen}(1988)}]{hist_reweighting1}
\bibinfo{author}{\bibfnamefont{A.~M.} \bibnamefont{Ferrenberg}}
  \bibnamefont{and} \bibinfo{author}{\bibfnamefont{R.~H.}
  \bibnamefont{Swendsen}}, \bibinfo{journal}{Phys. Rev. Lett.}
  \textbf{\bibinfo{volume}{61}}, \bibinfo{pages}{2635} (\bibinfo{year}{1988}).

\bibitem[{\citenamefont{Ferrenberg and Swendsen}(1989)}]{hist_reweighting2}
\bibinfo{author}{\bibfnamefont{A.~M.} \bibnamefont{Ferrenberg}}
  \bibnamefont{and} \bibinfo{author}{\bibfnamefont{R.~H.}
  \bibnamefont{Swendsen}}, \bibinfo{journal}{Phys. Rev. Lett.}
  \textbf{\bibinfo{volume}{63}}, \bibinfo{pages}{1195} (\bibinfo{year}{1989}).

\bibitem[{\citenamefont{Jolliffe}(1986)}]{PCA}
\bibinfo{author}{\bibfnamefont{I.~T.} \bibnamefont{Jolliffe}},
  \emph{\bibinfo{title}{Principal Component Analysis}}
  (\bibinfo{publisher}{Springer-Verlag}, \bibinfo{address}{New York, NY},
  \bibinfo{year}{1986}).

\bibitem[{\citenamefont{Smola et~al.}(2000)\citenamefont{Smola, Mangasarian,
  and Schoelkopf}}]{SparseKernel}
\bibinfo{author}{\bibfnamefont{A.~J.} \bibnamefont{Smola}},
  \bibinfo{author}{\bibfnamefont{O.~L.} \bibnamefont{Mangasarian}},
  \bibnamefont{and}
  \bibinfo{author}{\bibfnamefont{B.}~\bibnamefont{Schoelkopf}}, in
  \emph{\bibinfo{booktitle}{24th Annual Conference of Gesselschaft f{\"u}r
  Klassifikation}} (\bibinfo{organization}{University of Passau},
  \bibinfo{address}{Passau, Germany}, \bibinfo{year}{2000}),
  \bibinfo{note}{data Mining Institute Technical Reort 99-04}.

\bibitem[{\citenamefont{Ryckaert et~al.}(1977)\citenamefont{Ryckaert, Ciccotti,
  and Berendsen}}]{SHAKE}
\bibinfo{author}{\bibfnamefont{J.~P.} \bibnamefont{Ryckaert}},
  \bibinfo{author}{\bibfnamefont{G.}~\bibnamefont{Ciccotti}}, \bibnamefont{and}
  \bibinfo{author}{\bibfnamefont{H.}~\bibnamefont{Berendsen}},
  \bibinfo{journal}{J. Comp. Phys.} \textbf{\bibinfo{volume}{23}},
  \bibinfo{pages}{327} (\bibinfo{year}{1977}).

\bibitem[{\citenamefont{Gear et~al.}(2002)\citenamefont{Gear, Kevrekidis, and
  Theodoropoulos}}]{Theodoropoulos02}
\bibinfo{author}{\bibfnamefont{C.~W.} \bibnamefont{Gear}},
  \bibinfo{author}{\bibfnamefont{I.~G.} \bibnamefont{Kevrekidis}},
  \bibnamefont{and}
  \bibinfo{author}{\bibfnamefont{C.}~\bibnamefont{Theodoropoulos}},
  \bibinfo{journal}{Comput. Chem. Eng.} \textbf{\bibinfo{volume}{26}},
  \bibinfo{pages}{941} (\bibinfo{year}{2002}).

\bibitem[{\citenamefont{Gear}(2001)}]{GearDistr}
\bibinfo{author}{\bibfnamefont{C.~W.} \bibnamefont{Gear}}
  (\bibinfo{year}{2001}), \bibinfo{note}{technical Report NEC TR 2001-130}.

\bibitem[{\citenamefont{Kopelevich et~al.}(2004)\citenamefont{Kopelevich,
  Panagiotopoulos, and Kevrekidis}}]{KPK04b}
\bibinfo{author}{\bibfnamefont{D.~I.} \bibnamefont{Kopelevich}},
  \bibinfo{author}{\bibfnamefont{A.~Z.} \bibnamefont{Panagiotopoulos}},
  \bibnamefont{and} \bibinfo{author}{\bibfnamefont{I.~G.}
  \bibnamefont{Kevrekidis}}, \bibinfo{journal}{Submitted to J. Chem. Phys.}
  (\bibinfo{year}{2004}).

\bibitem[{\citenamefont{Torrie and Valleau}(1974)}]{Potential}
\bibinfo{author}{\bibfnamefont{G.~M.} \bibnamefont{Torrie}} \bibnamefont{and}
  \bibinfo{author}{\bibfnamefont{J.~P.} \bibnamefont{Valleau}},
  \bibinfo{journal}{Chem. Phys. Letters} \textbf{\bibinfo{volume}{28}},
  \bibinfo{pages}{578} (\bibinfo{year}{1974}).

\bibitem[{\citenamefont{Rico-Mart{\'i}nez
  et~al.}(2004)\citenamefont{Rico-Mart{\'i}nez, Gear, and Kevrekidis}}]{RMGK04}
\bibinfo{author}{\bibfnamefont{R.}~\bibnamefont{Rico-Mart{\'i}nez}},
  \bibinfo{author}{\bibfnamefont{C.~W.} \bibnamefont{Gear}}, \bibnamefont{and}
  \bibinfo{author}{\bibfnamefont{I.~G.} \bibnamefont{Kevrekidis}},
  \bibinfo{journal}{J. Comp. Phys.} \textbf{\bibinfo{volume}{196}},
  \bibinfo{pages}{474} (\bibinfo{year}{2004}).

\bibitem[{\citenamefont{Gillespie}(1976)}]{SA1}
\bibinfo{author}{\bibfnamefont{D.~T.} \bibnamefont{Gillespie}},
  \bibinfo{journal}{J. Comp. Phys.} \textbf{\bibinfo{volume}{22}},
  \bibinfo{pages}{403} (\bibinfo{year}{1976}).

\bibitem[{\citenamefont{Gillespie}(1977)}]{SA2}
\bibinfo{author}{\bibfnamefont{D.~T.} \bibnamefont{Gillespie}},
  \bibinfo{journal}{J. Phys. Chem.} \textbf{\bibinfo{volume}{81}},
  \bibinfo{pages}{2340} (\bibinfo{year}{1977}).

\bibitem[{\citenamefont{Gear}(2002)}]{Gear_Newton02}
\bibinfo{author}{\bibfnamefont{C.~W.} \bibnamefont{Gear}}
  (\bibinfo{year}{2002}), \bibinfo{note}{technical Report NECI TR 2002-025N,
  can be obtained as http://www.neci.nj.nec.com/homepages/cwg/steadystate.pdf}.

\bibitem[{\citenamefont{Drews et~al.}(2003)\citenamefont{Drews, Braatz, and
  Alkire}}]{Braatz03}
\bibinfo{author}{\bibfnamefont{T.~O.} \bibnamefont{Drews}},
  \bibinfo{author}{\bibfnamefont{R.~D.} \bibnamefont{Braatz}},
  \bibnamefont{and} \bibinfo{author}{\bibfnamefont{R.~C.}
  \bibnamefont{Alkire}}, \bibinfo{journal}{J. Electrochem. Soc.}
  \textbf{\bibinfo{volume}{150}}, \bibinfo{pages}{C807} (\bibinfo{year}{2003}).

\bibitem[{\citenamefont{Kelley}(1999)}]{Kelley_optimization_book}
\bibinfo{author}{\bibfnamefont{C.~T.} \bibnamefont{Kelley}},
  \emph{\bibinfo{title}{Iterative Methods for Optimization}}
  (\bibinfo{publisher}{SIAM}, \bibinfo{address}{Philadelphia},
  \bibinfo{year}{1999}).

\bibitem[{\citenamefont{Spall}(2000)}]{Spall00}
\bibinfo{author}{\bibfnamefont{J.~C.} \bibnamefont{Spall}},
  \bibinfo{journal}{IEEE Trans. Automat. Contr.} \textbf{\bibinfo{volume}{45}},
  \bibinfo{pages}{1839} (\bibinfo{year}{2000}).

\bibitem[{\citenamefont{Siettos et~al.}(2004)\citenamefont{Siettos, Kevrekidis,
  and Maroudas}}]{SMK04}
\bibinfo{author}{\bibfnamefont{C.~I.} \bibnamefont{Siettos}},
  \bibinfo{author}{\bibfnamefont{I.~G.} \bibnamefont{Kevrekidis}},
  \bibnamefont{and} \bibinfo{author}{\bibfnamefont{D.}~\bibnamefont{Maroudas}},
  \bibinfo{journal}{Int. J. Bifurcations and chaos}
  \textbf{\bibinfo{volume}{14}}, \bibinfo{pages}{207} (\bibinfo{year}{2004}).

\bibitem[{\citenamefont{Shaffer and Chakraborty}(1993)}]{Chakraborty93}
\bibinfo{author}{\bibfnamefont{J.~S.} \bibnamefont{Shaffer}} \bibnamefont{and}
  \bibinfo{author}{\bibfnamefont{A.~K.} \bibnamefont{Chakraborty}},
  \bibinfo{journal}{Macromolecules} \textbf{\bibinfo{volume}{26}},
  \bibinfo{pages}{1120} (\bibinfo{year}{1993}).

\bibitem[{\citenamefont{Kang and Weinberg}(1989)}]{Kang89}
\bibinfo{author}{\bibfnamefont{H.~C.} \bibnamefont{Kang}} \bibnamefont{and}
  \bibinfo{author}{\bibfnamefont{W.~H.} \bibnamefont{Weinberg}},
  \bibinfo{journal}{J. Chem. Phys.} \textbf{\bibinfo{volume}{90}},
  \bibinfo{pages}{2824} (\bibinfo{year}{1989}).

\bibitem[{\citenamefont{Snurr et~al.}(1994)\citenamefont{Snurr, Bell, and
  Theodorou}}]{SnurrTST}
\bibinfo{author}{\bibfnamefont{R.~Q.} \bibnamefont{Snurr}},
  \bibinfo{author}{\bibfnamefont{A.~T.} \bibnamefont{Bell}}, \bibnamefont{and}
  \bibinfo{author}{\bibfnamefont{D.~N.} \bibnamefont{Theodorou}},
  \bibinfo{journal}{J.~Phys. Chem.} \textbf{\bibinfo{volume}{98}},
  \bibinfo{pages}{11948} (\bibinfo{year}{1994}).

\bibitem[{\citenamefont{{\"O}ttinger}(1996)}]{Oettinger}
\bibinfo{author}{\bibfnamefont{H.~C.} \bibnamefont{{\"O}ttinger}},
  \emph{\bibinfo{title}{Stochastic Processes in Polymeric Fluids. Tools and
  Examples for Developing Simulation Algorithms}}
  (\bibinfo{publisher}{Sringer-Verlag}, \bibinfo{address}{Berlin-Heidelberg},
  \bibinfo{year}{1996}).

\bibitem[{\citenamefont{Melchior and {\"O}ttinger}(1995)}]{Oettinger1}
\bibinfo{author}{\bibfnamefont{M.}~\bibnamefont{Melchior}} \bibnamefont{and}
  \bibinfo{author}{\bibfnamefont{H.~C.} \bibnamefont{{\"O}ttinger}},
  \bibinfo{journal}{J. Chem. Phys} \textbf{\bibinfo{volume}{103}},
  \bibinfo{pages}{9506} (\bibinfo{year}{1995}).

\end{thebibliography}

\newpage


\end{document}